\documentclass[journal,comsoc,onecolumn,12pt]{IEEEtran}
\usepackage{setspace}
\usepackage{bm}
\usepackage{authblk}
\usepackage{cite}
\usepackage{graphicx}
\usepackage{amsmath}
\usepackage{array}
\usepackage{amssymb}
\usepackage{amsfonts}
\usepackage{epstopdf}
\usepackage{placeins}
\usepackage{caption}
\usepackage{algorithm, algorithmic}
\usepackage{soul}
\usepackage{multirow}

\newtheorem{lemma}{Lemma}

\newtheorem{remark}{Remark}

\newtheorem{theorem}{Theorem}

\newtheorem{example}{Example}

\usepackage{epsfig}
\usepackage{color}
\usepackage{psfrag}
\usepackage{circuitikz}
\usepackage{tikz}
\usepackage{balance}

\usepackage{mathtools}

\input{epsf.sty}

\begin{document}
\title{An Optimal Decentralized Multi-access Coded Caching System }

\author{
	\IEEEauthorblockN{Monolina Dutta,~Anoop Thomas,~\IEEEmembership{Member,~IEEE,} and B. Sundar Rajan,~\IEEEmembership{Life Fellow,~IEEE,}}
	\thanks{M. Dutta and A. Thomas are with the School of Electrical Sciences, Indian Institute of Technology Bhubaneswar,
		Khorda, 752050, India (email: md18@iitbbs.ac.in, anoopthomas@iitbbs.ac.in). B. S. Rajan is with the Dept. of Electrical Communication Engineering, Indian Institute of Science, Bengaluru, 560012, India (email: bsrajan@iisc.ac.in). }
}
\maketitle
\doublespacing

\begin{abstract}

In this paper, we consider a multi-access coded caching system with decentralized prefetching, where a server hosts $N$ files, each of size $F$ bits, and is connected to $K$ users through a shared link. There are $c$ caches distributed across the network and each of the $K$ users connects to a random set of $r\leq c$ caches. Initially, we consider the model in which each of the cache subsets is accessed by exactly a specific number of users. For this model, a novel linear delivery scheme is introduced, using which the closed-form expression for the per-user delivery rate is computed. Furthermore, using techniques from index coding, the optimality of the proposed linear delivery scheme among all linear delivery schemes is proved. The results of the decentralized shared caching and conventional decentralized caching schemes are recovered as special cases of the proposed model. The model is further generalized by allowing each cache subset to serve any number of users. This enhances the flexibility of the system, enabling it to accommodate any arbitrary number of users. A delivery scheme is proposed for the generalized model and is shown to be optimal for certain user-to-cache associations.


\end{abstract}
\begin{IEEEkeywords}
Coded Caching, decentralized caching, index coding, multi-access coded caching.
\end{IEEEkeywords}

\section{Intoduction}

{I}n this era of new-generation networks, internet and wireless networks are faced with the challenge of handling exponentially increasing data traffic due to the increasing use of multimedia sharing platforms. The nature of the distribution of such traffic shows strong temporal variability. As a result, the network encounters congestion during peak traffic hours while remains relatively under-utilized during off-peak hours. Coded caching is an effective strategy for mitigating network congestion, especially during peak hours by storing portions of data in the cache memories during off-peak hours \cite{MaN1}.


There are two phases in a caching problem: the prefetching phase and the delivery phase. The prefetching phase could either be centralized wherein a central server supervises the prefetching process \cite{MaN1}, or decentralized where the files are randomly cached by the users without the involvement of the central server \cite{maddah2014decentralized}. During the delivery phase, the demands of the users are revealed following which, the server devises transmissions such that the users are able to decode their demanded files from these transmissions and the cached contents.

Over recent years, there have been several reported extensions on the coded caching problem \cite{Combi_multi24,Combi_multi23, filesize, KNMDHeirarchical, onlinecaching, D2D, nonuniformdemands, Array, WUCCMF, shared}. The focus of this paper is on multi-access coded caching, which allows a user to access multiple caches. Specifically, each user can access a set of $r$ caches in the network, where $r$ is called the cache access degree. The multi-access coded caching was motivated by the heterogeneous cellular architecture for a 5G cellular network with multiple Access Points (APs) \cite{Main_Access}. Each AP can cache certain contents and has overlapping coverage areas, which enable each user to access multiple cache contents.

Various existing literature on multi-access coded caching largely focus on centralized prefetching.
In this setting, \cite{MAGains,firstmaJ,SasiImproved,CLWZC} have incorporated multi-access coded caching framework with cyclic wrap-around cache access, where each user is allowed to access a distinct set of consecutive caches. It is to be noted that in cyclic wrap-around cache access, the number of users is equal to the number of caches in the network. In order to make the system flexible in terms of the number of users in the network, a centralized multi-access coded caching system is considered in \cite{MaNAccess, Combi_multi24, Combi_multi23}, where each user can access an exclusive set of $r$ caches. In this framework, a set of $r$ caches is allowed to serve one user, which allows the number of users in the network to be $\binom{c}{r}$. Further, \cite{MALowerB} considers the same system model as in \cite{MaNAccess} and utilizes techniques from index coding to derive a lower bound on transmission rates and to prove the optimality of the delivery scheme in \cite{MaNAccess}. Furthermore, in another multi-access coded caching model under centralized setting, the cache access degree $r$ can vary across different users and each cache subset of cardinality $r$ is allowed to serve equal number of users \cite{Combi}.

Several studies in the literature have also investigated multi-access coded caching with decentralized prefetching. Decentralized prefetching in multi-access coded caching is desirable since the centralized schemes require the knowledge of the exact number of caches during the prefetching phase. 
The current network scenario encounters a large number of routers and access points which can act as caches. This makes it difficult to know the exact number of caches that take part in the delivery phase during the prefetching phase since it is highly improbable that all of them will be simultaneously active. A prior work in multi-access coded caching system with decentralized prefetching considers cyclic wrap-around cache access \cite{Main_Access}. Further, \cite{Main_Access} divides the demanded contents into discrete levels based on popularity and proposes a corresponding delivery algorithm. Additionally, the scheme in \cite{Conf_PMA} considers decentralized prefetching with cyclic wrap-around cache access and derives both upper and lower bounds on the per user delivery rate.  However, similar system models, considering the scenario in \cite{Combi} for a specific $r$ or relaxing the constraint on the number of users in the system model of \cite{Main_Access, Conf_PMA}, have not been considered in the decentralized prefetching setup which motivates us for this study.

To address the research gap outlined above, a multi-access coded caching model is considered in this paper wherein an arbitrary user is capable of accessing any set of $r$ caches instead of a distinct set of $r$ consecutive caches, as outlined in \cite{Conf_PMA}. Further, each set of $r$ caches is considered to serve $L$ users in contrary to the model outlined in \cite{Main_Access}, where each cache subset serves only one user. Here-forth, in this paper, such caching system is referred to as {\em decentralized multi-access coded caching system with cache access degree $r$ and user association $L$}. This framework is further extended to a more general setting, where each of these subset of $r$ caches is accessed by any arbitrary number of users. This generalization removes all constraints on the number of users in the network. Such caching system is referred to as {\em decentralized multi-access coded caching system with cache access degree $r$}. 

The novel contributions of this paper are summarized as follows:
\begin{itemize}
	\item A linear delivery scheme is proposed for the decentralized multi-access coded caching system with cache access degree $r$ and user association $L$, and a closed-form expression for the delivery rate of the proposed scheme is derived.
	\item The proposed scheme is demonstrated to be a generalization of the delivery schemes in \cite{maddah2014decentralized}, and \cite{MShared}. Specifically, the rate expressions in \cite{maddah2014decentralized} and \cite{MShared} are shown to be special cases of the proposed rate expression.
	\item Motivated by \cite{MShared, KVRDecentralized,ECSBP}, using techniques from index coding, the optimality of the proposed linear delivery scheme is proved among all linear delivery schemes for the decentralized multi-access coded caching setup with cache access degree $r$ and user association $L$.
 \item A linear delivery scheme is proposed for the decentralized multi-access coded caching system with cache access degree $r$. The optimality of the proposed delivery scheme for this general setting is proved for certain system parameters. 
\end{itemize}

The remainder of this paper is organized as follows. 
In Section \ref{SECTION:Scheme_1}, we present a system model for the multi-access coded caching and provide a review on index coding. Section \ref{delivery} provides a detailed description of the proposed linear delivery scheme for the considered framework along with the computation of transmission rate. 
The results pertaining to the optimality of the proposed delivery scheme are presented in Section \ref{optimal}. In Section \ref{ext}, a generalized multi-access coded caching model is proposed and the corresponding delivery scheme is provided. This is followed by concluding remarks in Section \ref{conclu}.

\textit{Notations:} For any positive integer $k$, $[k]$ denotes the set of integers $\{1,2,\ldots,k\}$. For a set $\mathcal{S}$, $| \mathcal{S}|$ denotes its cardinality. For any set $\mathcal{A}$, the set difference is represented as $\mathcal{S} \backslash \mathcal{A}$, where the elements in set $\mathcal{A}$ are removed from set $\mathcal{S}$. For any binary vector $\underline{b}$ of a certain length, the support of $\underline{b}$ denoted as $supp(\underline{b})$ is the set of coordinates corresponding to the non-zero entries in $\underline{b}$. 

\section{System Model and Preliminaries}
\label{SECTION:Scheme_1}
The client-server architecture considered in this paper comprises of a single server possessing $N$ files $W^1, W^2, \ldots, W^N$ each of size $F$ bits. The system comprises of $K$ users $1, 2, \ldots, K$ and $c$ caches $C_1, C_2, \ldots, C_c,$ each having normalized size $M$, with $0 \leq M \leq N$ capable of storing $MF$ bits. In multi-access coded caching, each user is allowed to access $r \leq c$ cache subsets, making $r$ the cache access degree. Consequently, there are in total $\binom{c}{r}$ subsets of cache indices with cardinality $r$. Each of these cache subsets is allowed to serve $L$ users. An instance of this system model is illustrated in Fig. \ref{fig1}, where $N=K = 12$, $c=4$, $L=2$, and $r=2$. 

A multi-access coded caching system with cache access degree $r$ and user association $L$ typically operates in three phases: the prefetching phase, the user-to-cache association phase, and the delivery phase. In the prefetching phase, decentralized prefetching, as outlined in \cite{maddah2014decentralized}, is utilized. Here, each cache independently stores a subset of $\frac{M F}{N}$ bits from each of the $N$ files, chosen uniformly at random. Accordingly, each bit of a file is equally likely of being cached in a cache with a probability of $\gamma = M/N$. Thus, we can say that each cache stores $\gamma F$ bits of each of the $N$ files at random. Considering $N$ files, the total size of the subfiles stored in one cache is given by $MF$ bits which is same as the size of each normalized cache memory. After the prefetching phase, each file $W^i$, $i\in [N]$ is viewed to be split into $W^i_{\mathcal{S}}, \forall \mathcal{S} \subseteq [c]$, where $W^i_{\mathcal{S}}$ denotes the set of bits of the file $W^i$ present in every cache having their indices in $\mathcal{S}$ and absent from all other caches. There are $2^c$ such subfiles per file. For sufficiently large file size $F$, by the law of large numbers, 
\begin{equation}
\label{size}
    \left|W^i_{\mathcal{S}}\right| \approx\gamma^{|\mathcal{S}|}(1-\gamma)^{c-|\mathcal{S}|} F.
\end{equation}
For the instance of the multi-access coded caching system considered in Fig. \ref{fig1}, the $N$ files are $W^1, W^2,$ $\ldots$, $W^{N}.$ Following the decentralized prefetching scheme, each file $W^i$ is viewed to be split into $2^4=16$ subfiles $ W^i_\mathcal{S}$, $\forall \mathcal{S} \subseteq \{1,2,3,4\}$.  

 After the prefetching phase in which the caches are filled, each user gets associated to a group of caches in the user-to-cache association phase. Specifically, each user is randomly allowed to access a set of $r$ caches, with the constraint that each cache subset can serve only $L$ users. Let $\mathfrak{C}_i, i \in \left[\binom{c}{r}\right]$ denote the subsets of cache indices each having cardinality $r$. These subsets of cache indices are arranged in a specific order by mapping each subset to a binary vector of length $c$. For each of these binary vectors, the numbering of the coordinates begins from the least significant bit. Consider an arbitrary subset of cache indices $\mathfrak{C}_z = \{{z_1}, {z_2}, \ldots, {z_{r}}\}$. The  set $\mathfrak{C}_z$ is mapped to a binary vector  $\underline{b}_z$ such that $supp(\underline{b}_z)$ is equal to $\mathfrak{C}_z$. The sets $\mathfrak{C}_1, \mathfrak{C}_2, \ldots, \mathfrak{C}_{\binom{c}{r}}$ are ordered such that the decimal equivalent of its corresponding mapped binary vectors are in ascending order.  Initially, we consider the multi-access coded caching model with each set of $r$ caches accessed by $L$ number of users. Since $L$ users access each set of $r$ caches, the decentralized multi-access coded caching system with cache access degree $r$ and user association $L$ can support a total of $K=L\binom{c}{r}$ users. The users associated to the set of cache indices $\mathfrak{C}_i$ is denoted as $\mathcal{U}_i$. The $\ell$-th user in the set $\mathcal{U}_i$ is indicated as $u_{i}(\ell)$, where $\ell\in [ L]$. The $K$ users are arranged as $u_{1}(1),u_{1}(2),\ldots,u_{1}(L), \ldots, u_{i}(\ell), \ldots, u_{\binom{c}{r}}(1), \ldots, u_{\binom{c}{r}}(L)$. In the decentralized multi-access coded caching system considered in  Fig. \ref{fig1}, each user set of $L=2$ users is associated with $r=2$ caches, resulting in an edge degree of two for each user set. The total number of cache-to-user edges is equal to $r \binom{c}{r} = 12$. As shown in Fig. \ref{fig1}, the users in the user sets are arranged as $u_{1}(1)=1$, $u_{1}(2)=2$, $\dots $, $u_{6}(2)=12$ .

 
\begin{figure}[t] \small
\begin{center}
\begin{tikzpicture}[scale=0.5]
\draw [, dashed] (4.75,12.75) rectangle  node {\scriptsize Server}  (6.5,12);
\draw  (8.5,12.75) rectangle  node {\small $W^2$} (11.25,12);
\draw  (8.5,13.5) rectangle  node {\small $W^1$} (11.25,12.75);
\draw  (8.5,10.5) rectangle  node {\small $W^N$} (11.25,9.75);
\draw  (6.75,7.25) circle (0.5cm) node {\small $C_1$} ;
\draw  (8.75,7.25) circle (0.5cm) node {\small $C_2$} ;
\draw  (11,7.25) circle (0.5cm) node {\small $C_3$} ;
\draw  (13,7.25) circle (0.5cm) node {\small $C_4$} ;
\draw [, dashed] (3,7.35) rectangle  node {\scriptsize Caches}  (4.75,6.75);
\draw  (8.5,12) rectangle  node {\normalsize $\vdots$} (11.25,10.5);
\draw  (4.8,4.5) rectangle  node {\scriptsize $\{1,2\}$} (6,3.75);
\draw  (6.3,4.5) rectangle  node {\scriptsize $\{3,4\}$ }(7.75,3.75);
\draw  (8.0,4.5) rectangle  node {\scriptsize $\{5,6\}$} (9.5,3.75);
\draw  (9.90,4.5) rectangle  node {\scriptsize $\{7,8\}$} (11.25,3.75);
\draw  (11.5,4.5) rectangle  node {\scriptsize $\{9,10\}$} (13,3.75);
\draw  (13.3,4.5) rectangle  node {\scriptsize $\{11,12\}$} (15.0,3.75);
\draw [short] (6.75,6.75) -- (5.5,4.5);
\draw [short] (5.5,4.5) -- (8.75,6.75);
\draw [short] (6.75,6.75) -- (7.25,4.5);
\draw [short] (7.25,4.5) -- (11,6.75);
\draw [short] (8.75,6.75) -- (9,4.5);
\draw [short] (9,4.5) -- (11,6.75);
\draw [short] (6.75,6.75) -- (10.75,4.5);
\draw [short] (10.75,4.5) -- (13,6.75);
\draw [short] (8.75,6.75) -- (12.5,4.5);
\draw [short] (12.5,4.5) --  (13,6.75);
\draw [short] (11,6.75) -- (14.25,4.5);
\draw [short] (14.25,4.5) -- (13,6.75);
\draw [short] (9.75,9.75) -- (6.75,7.75);
\draw [short] (9.75,9.75) -- (8.75,7.75);
\draw [short] (9.75,9.75) -- (11,7.75);
\draw [short] (9.75,9.75) -- (13,7.75);
\draw [, dashed] (2.0,4.35) rectangle  node {\scriptsize User Sets}  (4.0,3.75);
\end{tikzpicture}
\end{center}
\caption{System model of the multi-access coded caching scheme for $N=K=12$, $c=4$, $L=2$, and $r=2$.}
	\label{fig1}
\end{figure}
During the delivery phase, each user reveals its demand to the server. The demands of all the users in the system are described by a demand vector $\textbf{d} = \left(d(u_1(1)),\ldots, d(u_1(L)),\ldots, d(u_{\binom{c}{r}}(L))\right)$, where ${d}({u_{i}(\ell)})$ is the index of the file demanded by the user $u_{i}(\ell)$. The server has to ensure that each user can decode the demanded file from its transmissions and the cache contents of $r$ caches accessed by the user. A rate $R$ is said to be achievable if the server can deliver the demands of all the users without any error using $RF$ transmissions. Since the number of users served may be different for various schemes, rate per user or per user rate $\frac{R}{K}$ is the metric used for comparison between various schemes as in \cite{Katyal}, \cite{MA_PIR}. The focus of the paper is on the set of achievable per user rates considering distinct demands for the decentralized multi-access coded caching problem with cache access degree $r$ and user association $L$, which is denoted as $\mathcal{R}_D(M, r, L)$. The optimal achievable per user rate $R^{*}_D(M,r,L)$ is the minimum among $\mathcal{R}_D(M, r,L)$ achieved using linear delivery schemes. In this paper, we obtain this optimal rate $R^{*}_D(M,r,L)$.

We use techniques from index coding to prove the optimality of the proposed linear delivery scheme. In the remaining part of the section we review index coding and its connection with coded caching.

\subsection{Review of Index Coding}
The problem of index coding was introduced in \cite{index}. An instance $\mathcal{I}$ of an index coding problem  consists of a source possessing $n$ messages $x_1, x_2, \ldots, x_n$, and $k$ receivers $R_{1}, R_{2}, \ldots, R_{k}$. Each message $x_{i} \in \mathbb{F}_{q}, \forall i \in[n]$, where $\mathbb{F}_{q}$ is a finite field with cardinality $q$. A subset of messages with indices $\mathcal{X}_{i}$ is available with each receiver and is referred to as the side information. Each receiver $R_i$ demands a message with index $f(i)$ such that $f(i)\notin \mathcal{X}_{i}$. A linear index code is a set of linear combinations of messages transmitted by the source to deliver all the demands of the receivers. The minimum number of linear transmissions which the source has to make to deliver the demands is the length of an optimal linear index code denoted by $\kappa\left(\mathcal{I}\right)$. Another quantity of interest in an index coding problem $\mathcal{I}$ is the generalized independence number $\alpha(\mathcal{I})$. For every receiver $R_{i}$, $i\in[k]$, a set $\mathcal{B}_i\triangleq[n] \backslash\left(\{f(i)\} \cup \mathcal{X}_{i}\right)$ is considered. From sets $\mathcal{B}_i$, $i\in[k]$, the set $\mathcal{J}(\mathcal{I})\triangleq \cup_{i \in[k]}\left\{\{f(i)\} \cup B_{i}: B_{i} \subseteq\mathcal{B}_{i}\right\}$ is obtained. A subset $\mathcal{H} \subseteq [n]$ is called a generalized independent set of $\mathcal{I}$ if every subset of $\mathcal{H}$ belongs to the set $\mathcal{J}(\mathcal{I})$. The generalized independence number $\alpha(\mathcal{I})$ is the size of the largest generalized independent set of $\mathcal{I}$. In \cite{DSC}, it is shown that the generalized independence number is a lower bound on the length of an optimal linear index code. Hence for any index coding problem $\mathcal{I}$, we have 
\begin{equation}
\label{alphaaaa}
\alpha\left(\mathcal{I}\right) \leq \kappa\left(\mathcal{I}\right).
\end{equation}

\subsection{Index Coding and Coded Caching}
The delivery phase of a multi-access coded caching system can be modelled as an instance of an index coding problem. During the prefetching phase the cache contents are filled and these can be accessed by the users. These cached contents therefore act as side information in the corresponding index coding problem. Observe that there are $N^K$ possible demand vectors in a coded caching problem. Each of these $N^K$ possible demand vectors will result in a distinct index coding problem. Hence, a coded caching scheme consists of $N^K$ parallel index coding problems, each corresponding to one of the $N^K$ possible demand vectors. The work carried out in \cite{ECSBP, KVRDecentralized, firstma, MShared} and \cite{KSTR} have utilized the link between index coding and coded caching to obtain a lower bound on delivery rates. Following a similar idea, in this paper, a lower bound on transmission rates is derived, which is further utilized to prove the optimality of the proposed scheme.
\section{Delivery Scheme}
\label{delivery}
In this section, we propose an algorithm to deliver the demands of all the users for a decentralized multi-access coded caching system with cache access degree $r$ and user association $L$. The proposed algorithm uses a partitioning of subsets of cache indices with cardinality greater than or equal to $r$, for describing the transmissions made by the server. This description is required to prove the optimality of the proposed scheme in Section \ref{optimal} and further generalization in Section \ref{ext}. Hence, we first define certain sets and obtain a partition of subsets of cache indices with cardinality greater than or equal to $r$, before describing the algorithm. 

\subsection{Partitioning of Subsets of Cache Indices}
\label{Part}
For any $i\in\left[ \binom{c}{r}\right]$, we first define the set $\mathcal{C}_{i} \triangleq \bigcup\limits_{z\in [i] }\mathfrak{C}_z$. Let $\mathcal{P}_i$ denote the power set of $[c] \backslash \mathcal{C}_{i}$. We define $\mathcal{A}_i$ which is a collection of subsets of cache indices with cardinality greater than or equal to $r$ as 
\begin{equation}
\mathcal{A}_i \triangleq \left\{\mathfrak{C}_i\cup P_{(i,k)}: \forall P_{(i,k)}\in \mathcal{P}_i, k\in \big[|\mathcal{P}_i|\big]\right\}.
\end{equation}
There are $|\mathcal{P}_i|$ sets in $\mathcal{A}_i$. The $k$-th set in $\mathcal{A}_i$ is denoted as $A_{(i,k)}$ for $k\in \big[|\mathcal{P}_i|\big]$. Note that every set $A_{(i,k)}$ has cardinality greater than or equal to $r$ since it is $\mathfrak{C}_i \cup P_{(i,k)}$ and $|\mathfrak{C}_i | = r$. We show that the sets $\mathcal{A}_i, i \in \left[ \binom{c}{r}\right]$ forms a partition of subsets of cache indices with cardinality greater than or equal to $r$. 
\begin{lemma}
    \label{C}
    For any $k \in \left[\binom{c}{r}\right],$ the set $\mathcal{C}_{k} = [m],$ where $m = \underset{i\in \mathfrak{C}_{k}}{\max \,}\{i\}. $
\end{lemma}
\begin{IEEEproof}
    Recall that each $\mathfrak{C}_i$ is mapped to a binary vector $\underline{b}_i$, and these binary vectors $\underline{b}_1, \underline{b}_2,\dots, \underline{b}_{\binom{c}{r}},$ are such that the decimal equivalent of $\underline{b}_i$ is less than the decimal equivalent of $\underline{b}_j$ for $i<j$. Consider the set $\mathfrak{C}_k$ having $m = \underset{i\in \mathfrak{C}_{k}}{\max}\{i\}$. From the mapping of $\mathfrak{C}_i$ to the binary vectors, it can be observed that every $r$ element subset of $[m-1]$ will correspond to a $\mathfrak{C}_i$ with $i < k$. Hence, the set 
    $\mathcal{C}_{k} = \bigcup\limits_{z\in [k] }\mathfrak{C}_z$ has all the indices from $1$ to $m$. 
\end{IEEEproof}
    
  \begin{lemma}
\label{occurs}
 Any subset $\mathcal{S} \subseteq [c]$ having cardinality greater than or equal to $r$ belongs to the set $\underset{\forall i \in \left[\binom{c}{r}\right]}{\cup}\mathcal{A}_i$.
\end{lemma}
\begin{IEEEproof}     
       Consider $\mathcal{S}\subseteq [c]$ to be an arbitrary subset of cache indices  with $|\mathcal{S}|=s$ and $s\geq r$. There are $\binom{s}{r}$ subsets of set $\mathcal{S}$ each having cardinality $r$. Let these $\binom{s}{r}$ subsets be $\mathfrak{C}_{k_1},\mathfrak{C}_{k_2},\ldots,\mathfrak{C}_{k_{\binom{s}{r}}}$, with  $k_1\leq k_2\leq\ldots\leq k_{\binom{s}{r}}$. The set $\mathcal{S}$ can be viewed as  $\mathcal{S}=\mathfrak{C}_{k_1}\cup \mathcal{X}$, where $\mathfrak{C}_{k_1}\cap \mathcal{X}=\phi$. We now show that every element of $\mathcal{X}$ belongs to the set $[c]\backslash[m_{k_1}]$ where $m_{k_1} = \underset{i\in \mathfrak{C}_{k_1}}{\max \,}\{i\}$. Towards this, assume that there exists an element  $p\in\mathcal{X}$, such that $p < m_{k_1}$. Construct the set $\mathfrak{C}_{k_j} = \left(\mathfrak{C}_{k_1} \cup \{p\} \right) \setminus \{m_{k_1} \}$. Note that $|\mathfrak{C}_{k_j}| = r$ and $k_j < k_1$. This contradicts our assumption that $k_1\leq k_2\leq\ldots\leq k_{\binom{s}{r}}$. Hence, $\mathcal{X} \in [c]\backslash[m_{k_1}]$. This implies that  $\mathcal{X} \in \mathcal{P}_{k_1}$ from Lemma \ref{C}. Hence, $\mathcal{S} \in \mathcal{A}_{k_1}$. Since $\mathcal{S}$ is arbitrarily chosen, any subset of $[ c ]$ having cardinality greater than or equal to $r$ is also available in $\underset{\forall i \in \left[\binom{c}{r}\right]}{\cup}\mathcal{A}_i$.

    \end{IEEEproof}   

 \begin{lemma}
 \label{occurs_once}
For $\mathcal{A}_i$ and $\mathcal{A}_j$, where $i \neq j$ and $i,j\in \left[\binom{c}{r}\right]$, we have $\mathcal{A}_i\cap\mathcal{A}_j=\phi$.
\end{lemma}
\begin{IEEEproof}
     Without loss of generality, let us consider $i<j$ for the following proof. Let us assume that $\mathcal{A}_i\cap\mathcal{A}_j\neq\phi$ which implies that there exists atleast a set $\mathcal{S}\in \mathcal{A}_i\cap\mathcal{A}_j$ such that $\mathcal{S}=P_{(i,m)}\cup\mathfrak{C}_i$ and $\mathcal{S}=P_{(j,n)}\cup\mathfrak{C}_j$. Since, $\mathfrak{C}_i\neq \mathfrak{C}_j$, there exists atleast one $x$ such that $x\in\mathfrak{C}_i$ and $x\notin\mathfrak{C}_j$. For $P_{(i,m)}\cup\mathfrak{C}_i = P_{(j,n)}\cup\mathfrak{C}_j$ to be true, $x$ must belong to $P_{(j,n)}$. However, this is not possible as $P_{(j,n)} \subseteq [c]\backslash[m_j]$ (since $C_j=[m_j]$ from Lemma \ref{C}), $\mathfrak{C}_i\in [m_i]$, and $m_j>m_i$, where $m_j=\underset{z\in \mathfrak{C}_{j}}{\max}\{z\}$ and $m_i=\underset{z\in \mathfrak{C}_{i}}{\max}\{z\}$. This contradicts our assumption that $\mathcal{A}_i\cap\mathcal{A}_j\neq\phi$, which completes the proof.

\end{IEEEproof}

Using Lemmas \ref{occurs} and \ref{occurs_once}, we assert that every subset of $[c]$ with cardinality greater than or equal to $r$ occurs once in $\underset{\forall i \in \left[\binom{c}{r}\right]}{\cup}\mathcal{A}_i$. We now illustrate the above-defined sets through an example below. 

\renewcommand{\arraystretch}{1.6}
\begin{table}
	\begin{center}
    \caption{\centering{The set $\mathcal{P}_i$ and its corresponding $\mathcal{A}_i$,$\forall i\in\left[\binom{c}{r}\right]$ for Example \ref{Example:deliver_prep}.}}
        \begin{scriptsize}
			\begin{tabular}{|c|c|c|}
				\hline
				\textbf{$i$} & \textbf{$\mathcal{P}_i$} & \textbf{$\mathcal{A}_i$} \\
				\hline 
				1&$\{\{3,4\}, \{3\}, \{4\}, \phi \}$ & $\{\{1,2,3,4\}, \{1,2,4\}, \{1,2,3\}, \{1,2\}\}$\\
				\hline 
				2& $\{\{4\}, \phi \}$& $\{\{1,3,4\},\{1,3\}\}$\\
				\hline
				3&$\{\{4\}, \phi \}$ &$\{\{2,3,4\},\{2,3\}\}$\\
				\hline
				4&$\{\phi \}$& $\{\{1,4\}\}$\\
				\hline
				5&$\{\phi \}$&$\{\{2,4\}\}$\\
				\hline
				6&$\{\phi \}$&$\{\{3,4\}\}$\\
				\hline
			\end{tabular} 
        \end{scriptsize}
        \label{Table4}
	\end{center}
\end{table}
\begin{example}
	\label{Example:deliver_prep}
	Consider the decentralized multi-access coded caching system in Fig \ref{fig1}. There are $N$ files, $c=4$ number of caches, and cache access degree $r=2$. The set of cache indices $\{1,2,3,4\}$ is partitioned into $\binom{c}{r}=\binom{4}{2}=6$ subsets $\mathfrak{C}_1, \mathfrak{C}_2, \ldots, \mathfrak{C}_6$. The sets $\mathcal{P}_i$ and $\mathcal{A}_i$, $\forall i\in\left[\binom{c}{r}\right]$, are provided in Table \ref{Table4}.
	We can see from Table \ref{Table4} that  $\mathcal{A}_i, i \in  \forall i\in\left[\binom{c}{r}\right]$ partitions all subsets of cache indices having cardinality greater than or equal to two. We want to highlight that the sets $\mathcal{A}_i, i \in  \forall i\in\left[\binom{c}{r}\right]$ can be generated with the knowledge of cache access degree $r$ and the total number of caches $c$. The partitioning is independent of $L$ and the demands of the users. 
\end{example}

\subsection{Delivery Algorithm}
\label{delvi_algo}
In this subsection, we present our delivery scheme which utilizes the grouping of subsets of cache indices outlined in Section \ref{Part}. The transmissions performed by the server are described in Algorithm \ref{alg:placement}. For all $i\in \left[\binom{c}{r}\right]$, the server performs transmissions corresponding to every ${A}_{(i,k)}\in \mathcal{A}_i$, where $k\in \left[|\mathcal{A}_i|\right]$. For each ${A}_{(i,k)}$, the steps 5 through 7 in Algorithm 1 iterates $L$ times, implying that the server makes $L$ transmissions corresponding to the set ${A}_{(i,k)}$. The $\ell$-th transmission, $\ell\in  [L]$, is of the form $\underset{\substack{z \in \left[\binom{c}{r}\right]\\ \mathfrak{C}_z \subseteq {A}_{(i,k)}}}{\oplus} W^{{d}(u_{z}(\ell))}_{{A}_{(i,k)}  \backslash \mathfrak{C}_z}$, which is shown in step 6.
\begin{algorithm}
	\caption{Proposed algorithm}

	\begin{algorithmic}[1]
		\STATE \textbf{procedure} DELIVERY
		\FOR {$i=1, 2, \dots, \binom{c}{r}$} 
        \FOR {$k=1, 2, \dots, |\mathcal{A}_i|$}
		\FOR { ${A}_{(i,k)} \in \mathcal{A}_i$}
		\FOR {\textbf{all} $\ell\in  [L]$ }
		\STATE server transmits $\underset{\substack{z \in \left[\binom{c}{r}\right]\\ \mathfrak{C}_z \subseteq {A}_{(i,k)}}}{\oplus} W^{{d}(u_{z}(\ell))}_{{A}_{(i,k)}  \backslash \mathfrak{C}_z}$; 
		\ENDFOR
		\ENDFOR
		\ENDFOR
        \ENDFOR
		
		\STATE \textbf{end procedure}
	\end{algorithmic}
	\label{alg:placement}

\end{algorithm} 

In Algorithm \ref{alg:placement}, the transmissions are done corresponding to every ${A}_{(i,k)}\in \mathcal{A}_i$, where $k\in \left[|\mathcal{A}_i|\right]$. Recall that from Lemmas \ref{occurs} and \ref{occurs_once}, we assert that  $\underset{\forall i \in \left[\binom{c}{r}\right]}{\cup}\mathcal{A}_i$ correspond to all the subsets of $[c]$ having cardinality greater than or equal to $r$. Hence the server makes transmissions for every $\mathcal{S}\subseteq [c]$ having cardinality greater than or equal to $r$. For such a subset $\mathcal{S}$, the server makes $L$ transmissions. The $\ell$-th transmission is of the form $\underset{\substack{z \in \left[\binom{c}{r}\right]\\ \mathfrak{C}_z \subseteq \mathcal{S}}}{\oplus} W^{{d}(u_{z}(\ell))}_{\mathcal{S}  \backslash \mathfrak{C}_z}$, where $\ell\in [L]$. Hence, Algorithm \ref{alg:placement} can also be described by considering all such subsets $\mathcal{S}\subseteq [c]$ having cardinality greater than or equal to $r$. However, we have chosen to describe our delivery scheme as outlined in Algorithm \ref{alg:placement} so that the optimality of the proposed scheme can be proved in Section \ref{optimal} and our description could be extended to a more general setting in Section \ref{ext}.

\begin{example}
	\label{Example:AlgoDemo}
	Consider the instance of the decentralized multi-access coded caching setting described in Example \ref{Example:deliver_prep}. After the user-to-cache association phase, each of the $12$ users reveals its demands. Let the demands of these users be represented by a demand vector $\mathbf{d} = (1,2,3,4,5,6,7,8,9,10,11,12)$. The server makes transmissions corresponding to every subset of cache indices having cardinality greater than or equal to $r=2$. All such subsets are provided in Table \ref{Table4}. As an example, we explicitly illustrate the transmissions made for the set ${A}_{(1,3)}=\{1,2,3\}$ belonging to $\mathcal{A}_1$. The steps 5 through 7 of Algorithm 1 iterate $L=2$ times, implying that corresponding to the subset $\{1,2,3\}$, the server transmits twice. The first transmission as per step 6 of Algorithm 1 for $\ell=1$, is $W^{{d}(u_1(1))}_{\{3\}}\oplus W^{{d}(u_2(1))}_{\{2\}}\oplus W^{{d}(u_3(1))}_{\{1\}}.$ For the considered demand vector $\textbf{d}$, the above transmission is $W^1_{\{3\}}\oplus W^3_{\{2\}}\oplus W^5_{\{1\}}$. Similarly, the second transmission for $\ell=2$, is $W^{{d}(u_1(2))}_{\{3\}}\oplus W^{{d}(u_2(2))}_{\{2\}}\oplus W^{{d}(u_3(2))}_{\{1\}}$ which is same as $W^2_{\{3\}}\oplus W^4_{\{2\}}\oplus W^6_{\{1\}}$. All the transmissions made by the server in step $6$ of the algorithm, for each set in every cache partition $\mathcal{A}_i$, are explicitly provided in Table \ref{Table:Table1}. We now show how the demand of user $u_{1}(1)$ gets delivered using the proposed transmissions. User $u_{1}(1)$ requires the subfiles $W^1_{\phi}$, $W^1_{\{3\}}$, $W^1_{\{4\}}$ and $W^1_{\{3,4\}}$ of the demanded file $W^1$. The remaining subfiles of $W^1$ are available in the caches accessed by $u_{1}(1)$. The subfile $W^1_{\{3,4\}}$ is decoded from the XORed transmission corresponding to the subset $\{1,2,3,4\}$ since except $W^1_{\{3,4\}}$ all the files in the XORed transmission are available at the caches accessed by the user $u_{1}(1)$. The files $W^1_{\{3\}}$ and $W^1_{\{4\}}$ are decoded from the transmissions corresponding to the subsets $\{1,2,3\}$ and $\{1,2,4\}$ respectively. The subfile $W^1_{\phi}$ is directly obtained from the transmission corresponding to the subset $\{1,2\}$. Hence, the demand of the user $u_{1}(1)$ is delivered. Similarly, it can be verified using Table \ref{Table:Table1} that the demands of all the users are met using the transmissions from the server and the cache contents accessible to the respective users.
	\renewcommand{\arraystretch}{1.2}
	\begin{table}[t]
		\caption {\centering{Transmissions made by the server for Example \ref{Example:AlgoDemo}.}}
		\begin{center}
         \begin{scriptsize}  
			\begin{tabular}{|c|c|c|}
				\hline
				\multirow{1}{*}{\textbf{$\mathcal{A}_i$}}&\textbf{${A}_{(i,k)}$} & \textbf{Transmissions} \\
				\hline 
				\multirow{4}{*}{{$\mathcal{A}_1$}}&\{1,2,3,4\}&\!\!\!$W^1_{\{3,4\}}\!\oplus W^3_{\{2,4\}}\!\oplus W^5_{\{1,4\}}\!\oplus W^7_{\{2,3\}}\!\oplus W^9_{\{1,3\}}\!\oplus W^{11}_{\{1,2\}}$,\\& & $W^2_{\{3,4\}}\oplus W^4_{\{2,4\}}\oplus W^6_{\{1,4\}}\!\oplus W^8_{\{2,3\}}\!\oplus W^10_{\{1,3\}}\!\oplus W^12_{\{1,2\}}$\\ 
                \cline{2-3} 
                &\{1,2,4\}&$W^1_{\{4\}}\oplus W^7_{\{2\}}\oplus W^9_{\{1\}}$, $W^2_{\{4\}}\oplus W^8_{\{2\}}\oplus W^{10}_{\{1\}}$ \\
				\cline{2-3}  
				&\{1,2,3\}& $W^1_{\{3\}}\oplus W^3_{\{2\}}\oplus W^5_{\{1\}}$, $W^2_{\{3\}}\oplus W^4_{\{2\}}\oplus W^6_{\{1\}}$\\
                \cline{2-3} 
                &\{1,2\}&$W^1_{\phi}$, $W^2_{\phi}$\\
				\hline
				\multirow{3}{*}{{$\mathcal{A}_2$}}&\{1,3,4\}&$W^3_{\{4\}}\oplus W^7_{\{3\}}\oplus W^{11}_{\{1\}}$, $W^4_{\{4\}}\oplus W^8_{\{3\}}\oplus W^{12}_{\{1\}}$\\
				\cline{2-3}
                &\{1,3\}&$W^3_{\phi}$,$W^4_{\phi}$\\
                \hline
				\multirow{2}{*}{{$\mathcal{A}_3$}}&\{2,3,4\}&$W^5_{\{4\}}\oplus W^9_{\{3\}}\oplus W^{11}_{\{2\}}$, $W^6_{\{4\}}\oplus W^{10}_{\{3\}}\oplus W^{12}_{\{2\}}$\\
                \cline{2-3}   
                &\{2,3\}&$W^5_{\phi}$, $W^6_{\phi}$\\
				\hline
				\multirow{1}{*}{{$\mathcal{A}_4$}}&\{1,4\}&$W^7_{\phi},W^8_{\phi}$\\
				\hline
				\multirow{1}{*}{{$\mathcal{A}_5$}}&\{2,4\}&$W^9_{\phi},W^{10}_{\phi}$\\
				\hline
				\multirow{1}{*}{{$\mathcal{A}_6$}}&\{3,4\}&$W^{11}_{\phi},W^{12}_{\phi}$\\
				\hline
			\end{tabular} 
         \end{scriptsize}  
		\end{center}
		\label{Table:Table1}
	\end{table}
\end{example}

In Theorem \ref{lemma:delivery_scheme} below, we prove that the proposed algorithm delivers the demands of all the users in general.	
\begin{theorem}
    For a decentralized multi-access coded caching system with cache access degree $r$ and user association $L$, Algorithm 1 delivers the demands of all the users. 
	\label{lemma:delivery_scheme}
\end{theorem}
\begin{IEEEproof}
	Consider an arbitrary user $u_{a}(\ell)$ demanding a file $W^{{d}(u_{a}(\ell))}$, where $a\in\left[\binom{c}{r}\right]$. The bits in the demanded file $W^{{d}(u_{a}(\ell))}$  can be written as $\bigcup\limits_{\mathcal{D}\subseteq[c]} W^{{d}(u_{a}(\ell))}_{\mathcal{D}}$. All the subfiles $W^{{d}(u_{a}(\ell))}_{\mathcal{D}}$ with $\mathfrak{C}_a \cap \mathcal{D} \neq \phi$ are available with the caches associated to user $u_{a}(\ell)$. The server needs to deliver the bits $\bigcup\limits_{\mathcal{D}\subseteq[c]} W^{{d}(u_{a}(\ell))}_{\mathcal{D}}$, with $\mathfrak{C}_a \cap \mathcal{D} = \phi$. For any such set $\mathcal{D}_1$, where $\mathfrak{C}_a \cap \mathcal{D}_1 = \phi$, consider the set $\mathcal{S} = \mathcal{D}_1\cup \mathfrak{C}_a$. Since $\mathcal{S}\subseteq[c]$ has cardinality greater than or equal to $r$, from Lemmas \ref{occurs} and \ref{occurs_once}, we have $\mathcal{S}$ occurring exactly once in $\underset{\forall i \in \left[\binom{c}{r}\right]}{\cup}\mathcal{A}_i$. Accordingly, there exist exactly one $A_{(i,k)}$ that is mapped to $\mathcal{S}$, for some $i\in\left[\binom{c}{r}\right]$ and $k\in \left[|\mathcal{A}_i|\right]$. Thus, according to the proposed delivery scheme, the transmission $$\underset{\substack{z \in \left[\binom{c}{r}\right]\\ \mathfrak{C}_z \subseteq \mathcal{S}}}{\oplus} W^{{d}(u_{z}(\ell))}_{A_{(i,k)}  \backslash \mathfrak{C}_z}$$ is made by the server. Note that since  $\mathcal{D}_1=\mathcal{S}\backslash\mathfrak{C}_a=A_{(i,k)}\backslash\mathfrak{C}_a$, the above transmission includes a term $W^{{d}(u_{a}(\ell))}_{\mathcal{D}_1}$ XORed with other terms comprising the bits available in at least one cache having its index belonging to $\mathfrak{C}_a$. Thus, the user $u_{a}(\ell)$ is able to decode $W^{{d}(u_{a}(\ell))}_{\mathcal{D}_1}$ from the above transmission as all other XORed terms are available in its cache. Since $\mathcal{D}_1$ is arbitrarily chosen, any file $W^{{d}(u_{a}(\ell))}_{\mathcal{D}}$ with $\mathcal{C}_a \cap \mathcal{D} = \phi$ can be decoded by the user $u_{a}(\ell)$.  Furthermore, since the user $u_{a}(\ell)$ is arbitrarily chosen, the proposed scheme delivers all the demands of all the users. 
\end{IEEEproof}

\subsection{Per user transmission rate}
\label{PER_T_RATE}
For the proposed delivery scheme, Theorem \ref{lemma:rate} below computes the per user transmission rate.
\begin{theorem}
	For $N \geq K$, $r \in [c]$, and $0 \leq M \leq N$, the per user transmission rate of the delivery scheme described in Algorithm \ref{alg:placement} is
 \begin{equation}
 \label{rateeeee}
 \dfrac{L}{K}\sum_{s=r}^{c} \binom{c}{s}\gamma^{s-r}(1-\gamma)^{c-s+r}.
 \end{equation}
	\label{lemma:rate}
\end{theorem}
\begin{IEEEproof}
From Algorithm \ref{alg:placement}, corresponding to a set ${A}_{(i,k)} \in \mathcal{A}_i$, the server makes $L$ transmissions. The $\ell$-th transmission is of the form $$\underset{\substack{z \in \left[\binom{c}{r}\right]\\ \mathfrak{C}_z \subseteq \mathcal{S}}}{\oplus} W^{{d}(u_{z}(\ell))}_{A_{(i,k)}  \backslash \mathfrak{C}_z},$$ where $i\in\left[\binom{c}{r}\right]$, $k\in \left[|\mathcal{A}_i|\right]$, and $\ell \in[L]$. The size of each of these $L$ transmissions is $$\gamma^{|A_{(i,k)}|-r}(1-\gamma)^{c-|A_{(i,k)}|+r}F$$ bits. The server makes such transmissions for all ${A}_{(i,k)}$. As per Lemmas \ref{occurs} and \ref{occurs_once}, there is one-to-one mapping between the subsets of $[c]$ having cardinality greater than or equal to $r$ and the sets in $\underset{\forall i \in \left[\binom{c}{r}\right]}{\cup}\mathcal{A}_i$. This implies that the server makes transmissions for all subsets of $[c]$ having cardinality greater than or equal to $r$. Hence $|A_{(i,k)}|$ takes values from $r$ to $c$. For an arbitrary $s\geq r$, the number of $A_{(i,k)}$ having cardinality $s$ is $\binom{c}{s}$. The total number of  bits transmitted made by the server corresponding to $\binom{c}{s}$ subsets of $[c]$ 
having cardinality $s$ is $$L\binom{c}{s}\gamma^{s-r}(1-\gamma)^{c-s+r}F.$$  As $s$ ranges from $r$ to $c$, the total number of transmitted bits is $$L\sum_{s=r}^{c}\binom{c}{s}\gamma^{s-r}(1-\gamma)^{c-s+r}F.$$ Upon normalizing over $F$ and $K$, we obtain the per user transmission rate. This completes the proof.

\end{IEEEproof}

\begin{remark}
\label{remark2}
	The problem considered in \cite{MShared} for the scenario of each cache being accessed by an equal number of users is a special case of our proposed model with $r=1$. Observe that when $r=1$, the per user rate expression in \eqref{rateeeee} reduces to
\begin{eqnarray}
\label{shared_2_rates}
\dfrac{L}{K}\sum_{s=1}^{c}\binom{c}{s}\gamma^{s-1}(1-\gamma)^{c-s+1}\! \! \! \!&=&\! \! \! \!\! \!\dfrac{L(1-\gamma)}{K\gamma}\sum_{s=1}^{c}\binom{c}{s}\gamma^{s}(1-\gamma)^{c-s}\nonumber\\
&\stackrel{(a)}{=}& \! \! \! \!
\dfrac{L\left(1-\gamma \right)}{K\gamma}\left[1-(1-\gamma)^c\right] \nonumber\\
&\stackrel{(b)}{=}& \! \! \! \! \dfrac{L}{K}\sum_{i=1}^{c}\left(1-\gamma\right)^i.
\end{eqnarray}
The step $(b)$ in \eqref{shared_2_rates} is obtained by viewing $\dfrac{L\left(1-\gamma\right)}{K\gamma}\left[1-(1-\gamma)^c\right]$ as the sum of geometric progression $\dfrac{L}{K}\sum_{i=1}^{c}\left(1-\gamma\right)^i$. The total transmission rate $L\sum_{i=1}^{c}\left(1-\gamma\right)^i$ is same as that presented in \cite{MShared} for the decentralized shared caching scenario with each cache serving equal number of users. 
On further substituting $L=1$, $\forall i\in[c]$ in step $(a)$ of \eqref{shared_2_rates}, the total rate $$\frac{1}{\gamma}(1-\gamma)(1 - (1-\gamma)^{c})$$ is obtained, which is the rate expression derived in \cite{maddah2014decentralized} for the classical decentralized coded caching.
\end{remark}

\begin{figure}[h!]
		\begin{center}
		\includegraphics[scale=0.65]{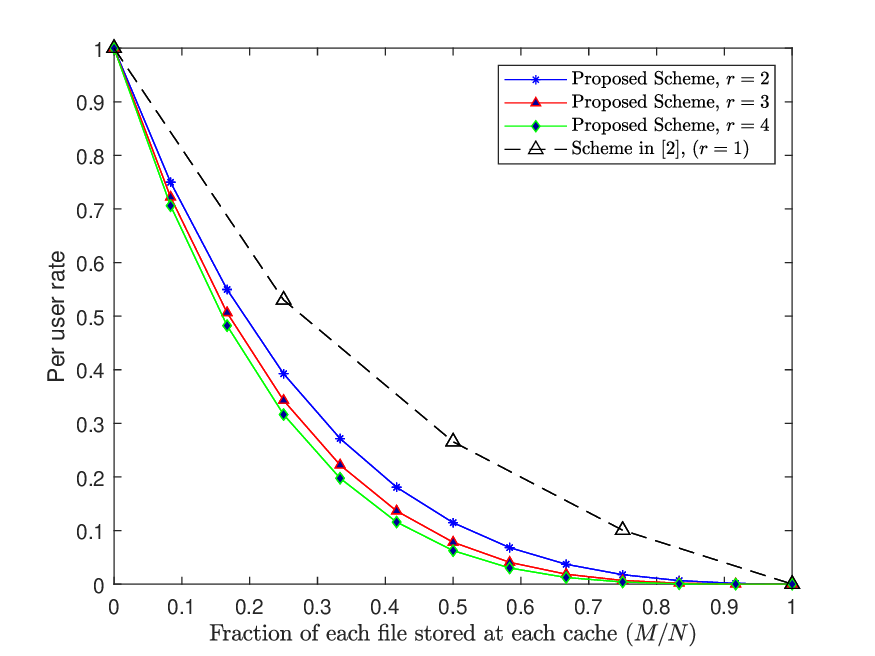}
		\caption{Per user transmission rates of our proposed scheme $\forall r\in \{2,3,4\}$ considering $c=4$ and $N=K=12$, and the per user transmission rate in \cite{maddah2014decentralized} for $c=4$ and $N=K=4$ are plotted against $\frac{M}{N}$.}
		\label{fig10}
  \end{center}
	\end{figure}

The per user transmission rates of our proposed scheme $\forall r\in \{2,3,4\}$ considering $c=4$ and $N=K=12$, and the per user transmission rate presented in \cite{maddah2014decentralized} for $c=4$ and $N=K=4$ are illustrated in Fig. \ref{fig10}. It is observed that the per user transmission rates reduce with the increase in the cache access degree $r$ as expected.
 
 \begin{remark}
 \label{rate_2_explain}
     An alternate expression for the per user transmission rate, obtained in Theorem \ref{lemma:rate}, is computed using the delivery scheme outlined in Algorithm \ref{alg:placement}. Recall that for an arbitrary set $A_{(i,k)} \in \mathcal{A}_i$, $\forall i\in\left[\binom{c}{r}\right]$, the server makes $L$ transmissions, each of size $\gamma^{|A_{(i,k)}|-r}(1-\gamma)^{c-|A_{(i,k)}|+r}F$ bits. Therefore, the total number of bits transmitted corresponding to $\underset{\forall i \in \left[\binom{c}{r}\right]}{\cup}\mathcal{A}_i$ is 
\begin{equation}
L\sum_{i=1}^{\binom{c}{r}}\sum_{k=1}^{|\mathcal{A}_i|}\gamma^{|A_{(i,k)}|-r}(1-\gamma)^{c-|A_{(i,k)}|+r}F.
\label{partition_rate}
\end{equation}
Upon normalizing over $F$ and $K$, we get per user transmission rate. From Lemmas \ref{occurs} and \ref{occurs_once}, we can say that the rate obtained in \eqref{partition_rate} is same as that computed in \eqref{rateeeee}. This alternate expression for the per user transmission rate obtained above is utilized in Sections \ref{optimal} and \ref{ext}.
 \end{remark}
\subsection{Comparison with centralized scheme}
\begin{figure}[h!]
		\centering	
		\includegraphics[scale=0.65]{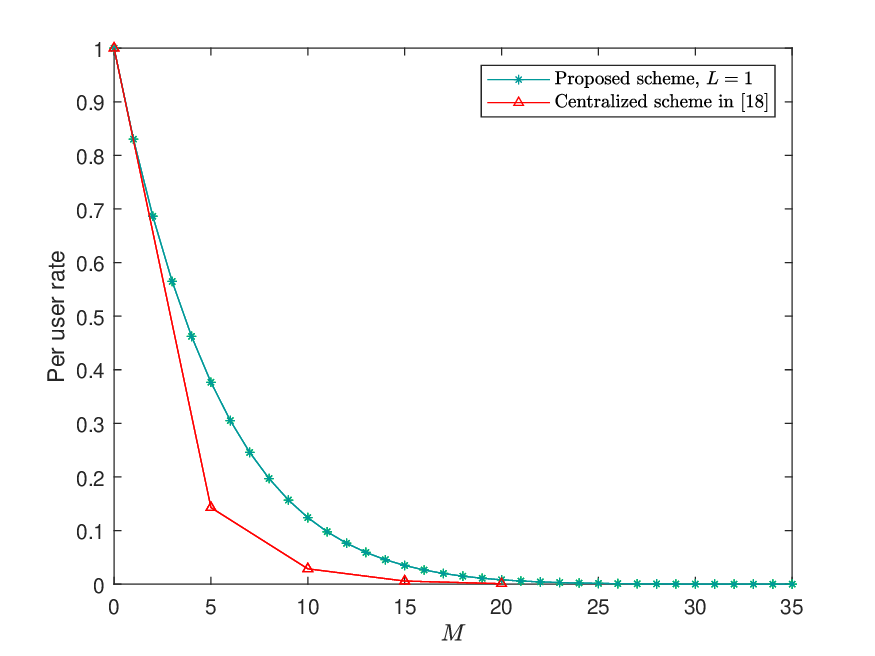}
		\caption{Per user transmission rates of our proposed scheme and the scheme in \cite{MaNAccess} considering $c=7$, $r=3$, and $N=K=35$ are plotted against $M$.}
		\label{fig100}
	\end{figure}

    For $L=1$, the proposed system model in this paper becomes the model considered in \cite{MaNAccess}. However, this paper considers decentralied prefetching whereas the prefetching is centralized in \cite{MaNAccess}. We compare the per user transmission rate for both the schemes considering an instance of the caching problem having $c=7$, $r=3$, and $N=K=35$ in Fig. \ref{fig100}. As expected centralized prefetching outperforms decentralized prefetching. Nevertheless, decentralized prefetching is more preferable, especially in large networks.
\section{Optimality of the Delivery Scheme}
\label{optimal}
In this section, we prove the optimality of the proposed delivery scheme by using techniques from index coding. As mentioned in Section II, the delivery phase of every coded caching problem can be viewed as an instance of an index coding problem. For the decentralized multi-access coded caching with cache access degree $r$ and user association $L$, let the delivery phase correspond to an instance of an index coding problem $\mathcal{I}_{DM}$. Each bit of a file of size $F$ bits in the multi-access caching system corresponds to a distinct message in $\mathcal{I}_{DM}$. Each user in a decentralized multi-access coded caching system is equivalent to a collection of maximum of $F$ receivers, each demanding a single bit of the message, in $\mathcal{I}_{DM}$. Hence, the index coding problem $\mathcal{I}_{DM}$ has a maximum of $NF$ messages and $KF$ receivers. The contents cached during the prefetching phase behave as the side information. Every receiver of the index coding problem can be denoted by $u_{i}(\ell,y)$ which demands the $y$-th bit of the file demanded by the user $u_{i}(\ell)$ of the multi-access decentralized caching system, where $y\in [ F ]$, $\ell\in [ L] $ and $i \in \left[ \binom{c}{r}\right]$. For the receiver $u_{i}(\ell,y)$, let $\mathcal{X}_{i}$ denote the set of bits in $\mathfrak{C}_i$ and ${d}({u_{i}(\ell,y)})$ denote the demanded bit. We define the set $\mathcal{B}_{u_{i}(\ell,y)} = [ NF] \backslash\left(\{{d}({u_{i}(\ell,y)})\} \cup \mathcal{X}_{i}\right)$, using which we define the set $\mathcal{J}(\mathcal{I}_{DM})$ as $\mathcal{J}(\mathcal{I}_{DM}) =  \underset{i \in\left[ \binom{c}{r}\right]}{\cup}\underset{\ell\in [ L]}{\cup}\underset{y \in[ F]}{\cup}\left\{\{{d}({u_{i}(\ell,y)})\} \cup B_{u_{i}(\ell,y)}: B_{u_{i}(\ell,y)} \subseteq \mathcal{B}_{u_{i}(\ell,y)}\right\}$. We aim to determine the generalized independence number $\alpha\left(\mathcal{I}_{DM}\right)$. Consequently, we give an explicit construction of an independent set. Recall that $\mathcal{C}_{i}=\bigcup\limits_{z\in [i] }\mathfrak{C}_z$, $\forall i\in\left[ \binom{c}{r}\right]$ and $\mathcal{P}_{i}$ denote the power set of $[c] \setminus\mathcal{C}_{i}$, $\forall i\in\left[\binom{c}{r}\right]$. This grouping of sets assists in the construction of an independent set. 

A lower bound on the generalized independence number $\alpha\left(\mathcal{I}_{DM}\right)$ is obtained in Lemma \ref{thm:alpha1} below.
\begin{lemma}
	For the index coding problem $\mathcal{I}_{DM}$ corresponding to the delivery phase of a decentralized multi-access coded caching system with cache access degree $r$ and user association $L$, we have
    \begin{equation}
    \label{alpha_bound}
	\alpha\left(\mathcal{I}_{DM}\right)\geq L\left[ \sum_{i=1}^{\binom{c}{r}}\sum_{k=1}^{|\mathcal{P}_i|}\gamma^{|P_{(i,k)}|}(1-\gamma)^{c-|P_{(i,k)}|} \right]F.
    \end{equation}
	\label{thm:alpha1}
\end{lemma}
\begin{IEEEproof}
	This statement is proved by explicitly constructing an independent set of $\mathcal{I}_{DM}$. We consider the set of bits 
\begin{equation}
\label{independent}
 \mathcal{Y}(\mathcal{I}_{DM})=\bigcup\limits_{i\in\left[ \binom{c}{r}\right],\ell\in[L] ,\mathcal{S}^{\prime} \in \mathcal{P}_{i} } W^{u_{i}(\ell)}_{\mathcal{S}^{\prime}}
 \end{equation}
 and subsequently show that the set $\mathcal{Y}(\mathcal{I}_{DM})$ is an independent set by proving that every subset of $\mathcal{Y}(\mathcal{I}_{DM})$ belongs to the set $\mathcal{J}(\mathcal{I}_{DM})$. Since each set of bits in the set $\mathcal{Y}(\mathcal{I}_{DM})$ is demanded by a set of receivers, this confirms that these set of bits are present as singletons in the set $\mathcal{J}(\mathcal{I}_{DM})$. Consider the set of bits in any $J$ subfiles in $\mathcal{Y}(\mathcal{I}_{DM})$ as $$D=\left\{W^{u_{z_{1}}(\ell_1)}_ {\mathcal{S}_{1}}, W^{u_{z_{2}}(\ell_2)}_{\mathcal{S}_{2}}, \ldots, W^{u_{z_{J}}(\ell_J)}_{ \mathcal{S}_{J}}\right\} \subseteq \mathcal{Y}(\mathcal{I}_{DM}),$$ where ${z}_{1} \leq {z}_{2} \leq \ldots \leq z_{J}$, $\mathcal{S}_{j}\in \mathcal{P}_{z_{j}}$, $z_{j}\in[\binom{c}{r}]$, $\ell_j \in [L] $, and $j\in[J]$. To show that the set $D$ belongs to $\mathcal{J}(\mathcal{I}_{DM})$, it is enough to show that the set of receivers, each demanding a bit in $W^{u_{z_{1}}(\ell_1)}_{ \mathcal{S}_{1}} $, which are associated to the user $u_{z_1}(\ell_1)$ do not have any bits in $D$ as their side information. To prove this, let us consider an arbitrary $z_j$ such that $z_j\geq z_1$. Recall that $\mathcal{C}_{z_j} =  \bigcup\limits_{p\in [z_j] }\mathfrak{C}_p$ and $\mathcal{P}_{z_{j}}$ is the power set of $[c]\backslash \mathcal{C}_{z_j}$. This implies that $\mathfrak{C}_{z_1}\cap\mathcal{S}_j=\phi$ as $\mathcal{S}_{j}\in \mathcal{P}_{z_{j}}$ and $z_j\geq z_1$. Since $z_j$ is arbitrarily chosen and ${z}_{1} \leq {z}_{2} \leq \ldots \leq z_{J}$, the set of receivers, each demanding a bit in $W^{u_{z_{1}}(\ell_1)}_{ \mathcal{S}_{1}} $, which are associated to the user $u_{z_1}(\ell_1)$ do not have any bits in $D$ as their side information. This implies that $D\in\mathcal{J}(\mathcal{I}_{DM})$. Since $D$ is arbitrarily chosen, this proves that all the subsets of $\mathcal{Y}(\mathcal{I}_{DM})$ are in the set $\mathcal{J}(\mathcal{I}_{DM})$.
	
	We now calculate the cardinality of the set $\mathcal{Y}(\mathcal{I}_{DM})$. Consider the set $\mathcal{P}_{i}$. For the $k$-th set $P_{(i,k)}$, the number of bits in $W^{u_{i}(\ell)}_{P_{(i,k)}}$ is equal to $$\gamma^{|P_{(i,k)}|}(1-\gamma)^{c-|P_{(i,k)}|}F.$$ Considering all sets in $\mathcal{P}_{i}$ we obtain the number of bits for a particular $\ell$ and $i$ as  $$\sum_{k=1}^{|\mathcal{P}_i|} \gamma^{|P_{(i,k)}|}(1-\gamma)^{c-|P_{(i,k)}|}F.$$ Since, $\ell \in [ L ]$ and $ i \in [ \binom{c}{r} ]$, the cardinality of $\mathcal{Y}(\mathcal{I}_{DM})$ is equal to 
	$$\left[ \sum_{i=1}^{\binom{c}{r}}L\sum_{k=1}^{|\mathcal{P}_i|}\gamma^{|P_{(i,k)}|}(1-\gamma)^{c-|P_{(i,k)}|} \right]F.$$ Since $\mathcal{Y}(\mathcal{I}_{DM})$ is a generalized independent set and $\alpha\left(\mathcal{I}_{DM}\right)$ is the size of the largest generalized independent set, the statement of this lemma is proved.
 \end{IEEEproof}
\begin{lemma}
\label{kappa_eq}
For the index coding problem $\mathcal{I}_{DM}$ corresponding to the delivery phase of a decentralized multi-access coded caching system with cache access degree $r$ and user association $L$, we have
\begin{equation}
 \label{kappa_eq_alpha}
 \kappa\left(\mathcal{I}_{DM}\right)\!=\!\!\left[L \sum_{i=1}^{\binom{c}{r}}\sum_{k=1}^{|\mathcal{A}_i|}\gamma^{|A_{(i,k)}|-r}(1-\gamma)^{c-|A_{(i,k)}|+r} \right]F.
 \end{equation}
\end{lemma}
 \begin{IEEEproof}
 For the index coding problem $\mathcal{I}_{DM}$, we have $\kappa\left(\mathcal{I}_{DM}\right)$ as the optimal solution. Since our delivery phase corresponds to an instance of $\mathcal{I}_{DM}$, the number of bits transmitted by our proposed scheme is a solution to this index coding problem. As $\kappa\left(\mathcal{I}_{DM}\right)$ represents the best possible solution, the total number of bits transmitted by our proposed delivery scheme serves as an upper bound on $\kappa\left(\mathcal{I}_{DM}\right)$. Therefore, utilizing \eqref{partition_rate}, we get $\kappa\left(\mathcal{I}_{DM}\right)\leq \left[ L\sum_{i=1}^{\binom{c}{r}}\sum_{k=1}^{|\mathcal{A}_i|}\gamma^{|A_{(i,k)}|-r}(1-\gamma)^{c-|A_{(i,k)}|+r} \right]F$. From the definition of the sets $\mathcal{P}_i$ and $\mathcal{A}_i$, observe that $|\mathcal{P}_i|=|\mathcal{A}_i|$ and $|P_{(i,k)}|=|A_{(i,k)}|-r$. Substituting these in \eqref{alpha_bound}, we get $\alpha\left(\mathcal{I}_{DM}\right)\geq \left[ L\sum_{i=1}^{\binom{c}{r}}\sum_{k=1}^{|\mathcal{A}_i|}\gamma^{|A_{(i,k)}|-r}(1-\gamma)^{c-|A_{(i,k)}|+r} \right]F$. Utilizing the relation $\alpha\left(\mathcal{I}_{DM}\right) \leq \kappa\left(\mathcal{I}_{DM}\right) $ from \eqref{alphaaaa}, we conclude that the upper and lower bounds on $\kappa\left(\mathcal{I}_{DM}\right)$ are tight. This completes the proof.

\end{IEEEproof}

In the theorem below, we obtain the optimal per user transmission rate for the proposed multi-access caching model.
\begin{theorem}
	For $r \in [ c ]$, $N \geq K$ and $0 \leq M \leq N$, the optimal per user transmission rate of the decentralized multi-access coded caching with cache access degree $r$ and user association $L$ is equal to \begin{equation}R^{*}_{{D}}(M,r)=\dfrac{L}{K}\sum_{s=r}^{c} \binom{c}{s}\gamma^{s-r}(1-\gamma)^{c-s+r}. \end{equation} 
	\label{thm:lowerbound}
\end{theorem}
\begin{IEEEproof}
Note that any delivery scheme transmits a minimum of $\kappa\left(\mathcal{I}_{DM}\right)$ bits to deliver all the demands.
	Using Lemma \ref{kappa_eq}, the minimum transmitted bits required to satisfy all the demands is $\kappa\left(\mathcal{I}_{DM}\right) = \left[ L\sum_{i=1}^{\binom{c}{r}}\sum_{k=1}^{|\mathcal{A}_i|}\gamma^{|A_{(i,k)}|-r}(1-\gamma)^{c-|A_{(i,k)}|+r} \right]F.$ Upon normalizing over $F$ and $K$, it exactly matches with \eqref{partition_rate}. As outlined in Remark \ref{rate_2_explain}, \eqref{partition_rate} is an alternate expression for \eqref{rateeeee}. Hence, the per user transmission rate of the proposed scheme is optimal among all possible linear delivery schemes for multi-access coded caching system with cache access degree $r$ and user association $L$. This completes the proof.
\end{IEEEproof} 

Further, we present an example to elucidate the generalized independent set $\mathcal{Y}(\mathcal{I}_{DM})$ along with Theorem \ref{thm:lowerbound}.
\begin{example}
	\label{Example:LB}
	Consider the decentralized multi-access coded caching setup of Example. \ref{Example:deliver_prep}. Following \eqref{independent} we construct the set $\mathcal{Y}(\mathcal{I}_{DM})$ as the union of the sets of bits $W^{u_i(\ell)}_{\mathcal{S}}$ for every $ \mathcal{S} \in \mathcal{P}_{i} $, $\forall i\in [6]$ and $\ell \in [ L ]$. All such $W^{u_i(\ell)}_{\mathcal{S}}$ are provided in Table \ref{Table:Table2}. We obtain $|\mathcal{Y}(\mathcal{I}_{DM})|$ as $\left[12(1-\gamma)^{4}+8\gamma(1-\gamma)^{3}+2\gamma^{2}(1-\gamma)^{2}\right]F,$ which reduces to $\left[2(1-\gamma)^{2}+4(1-\gamma)^{3}+ 6(1-\gamma)^{4}\right]F.$ Upon normalizing over $F$ and averaging over $K=12$ users, it matches with the per user transmission rate $\dfrac{1}{6}\sum_{s=2}^{4} \binom{4}{s}\gamma^{s-2}(1-\gamma)^{6-s}=\dfrac{1}{6}(1-\gamma)^{2}+ \dfrac{1}{3}\gamma(1-\gamma)^{3}+\dfrac{1}{2}(1-\gamma)^{4}.$ This is also illustrated in Fig. \ref{fig2}. Additionally, in Fig. \ref{fig2} the per user delivery rates and the corresponding lower bounds are shown to be matching for remaining $r\in\{1,2,3,4\}$ and corresponding $L=\frac{12}{\binom{4}{r}}$.
	\renewcommand{\arraystretch}{1.2}
	\begin{table}
 \caption{\centering{The sets of bits $\mathcal{Y}(\mathcal{I}_{DM})$ for $c=4, r=2$, and $L=2$, $\forall i\in \{1,2,\ldots,6\} $.}}
		\begin{center} 
  \begin{small}
			\begin{tabular}{|c|c|c|}
				\hline
				$i$ &\text{$\mathcal{P}_{i}$} & \text{$\mathcal{Y}(\mathcal{I}_{DM})$} \\
				\hline 
				1&\{3,4\}, \{3\}, \{4\}, $\phi$ &$W^1_{\{3,4\}}, W^2_{\{3,4\}}, W^1_{\{3\}}, W^2_{\{3\}}$,\\& & $W^1_{\{4\}}, W^2_{\{4\}}, W^1_\phi, W^2_\phi$\\
				\hline 
				2&\{4\},$\phi$&$W^3_{\{4\}}, W^4_{\{4\}}, W^3_\phi, W^4_\phi$\\
				\hline
				3&$\{4\},\phi$&$W^5_4,W^5_\phi, W^6_4,W^6_\phi$\\
				\hline
				4&$\phi$&$W^7_\phi,W^8_\phi$\\
				\hline
				5&$\phi$&$W^9_\phi,W^{10}_\phi$\\
				\hline
				6&$\phi$&$W^{11}_\phi,W^{12}_\phi$\\
				\hline
			\end{tabular} 
    \end{small}
		\end{center}
		\label{Table:Table2}
	\end{table}
	\begin{figure}[h!]
		\centering	
		\includegraphics[scale=0.65]{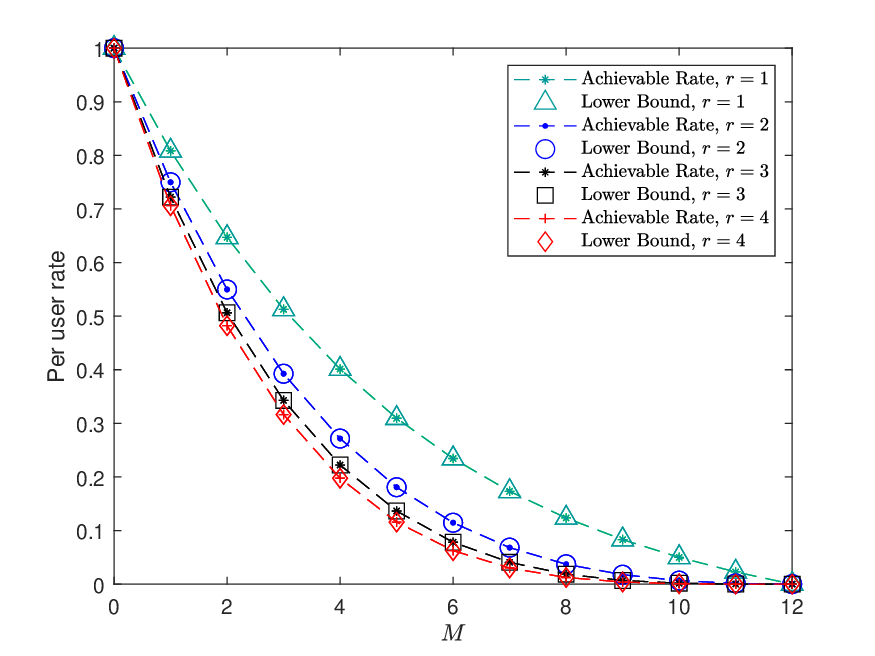}
		\caption{Proposed per user transmission rate and lower bound for the decentralized multi-access coded caching system with $N=K=12$ and $c=4$ for all $r\in\{1,2,3,4\}$, versus $M$.}
		\label{fig2}
	\end{figure}
\end{example}

\section{Extension to a more general setting}
\label{ext}
A restriction of the system model considered in Section \ref{SECTION:Scheme_1} is that each subset of $r$ caches is allowed to serve equal number of users thereby, fixing the number of users in the network to $L\binom{c}{r}$. We now extend this idea to a more generalized setting wherein each subset of $r$ caches is capable of serving any arbitrary number of users. This is accomplished by replacing $L$ with a user to cache association profile vector $\textbf{L} = (L_1, L_2, \ldots,{L}_{\binom{c}{r}}),$ where $L_i$ is the number of users associated to the cache subset $\mathfrak{C}_i$. Hence, the total number of users is $K=\sum_{i=1}^{\binom{c}{r}} {L}_i$. This generalized system model, also referred to as the decentralized multi-access coded caching system with cache access degree $r$, is novel and to the best of our knowledge has not been studied before in either centralized or decentralized settings.

\subsection{Delivery Scheme}
\label{del_gen}
In this subsection, we present a delivery scheme for the proposed decentralized multi-access coded caching system with cache access degree $r$ by modifying  Algorithm \ref{alg:placement} outlined in Section \ref{delvi_algo}. Recall that in Algorithm \ref{alg:placement}, the server makes $L$ transmissions for every $A(i,k), i \in \left[\binom{c}{r}\right], k \in \left[|\mathcal{A}_i|\right]$. This is possible because all subsets of $A(i,k)$ having cardinality $r$ are accessed by the same number of users. However, this is no longer true for the generalized setting. Hence, for each $A(i,k)$ the server has to make ${L}^{*}_{{A}_{(i,k)}}=\underset{\substack{\mathfrak{C}_z\subseteq {A}_{(i,k)}\\ z \in \left[\binom{c}{r}\right]}}{\text{max}}({L}_{z} )$ transmissions. This modification is integrated in steps $4-8$ of Algorithm \ref{algo2}.

\begin{algorithm}
	\caption{Proposed algorithm}

	\begin{algorithmic}[1]
		\STATE \textbf{procedure} DELIVERY
		\FOR {$i=1, 2, \dots, \binom{c}{r}$} 
        \FOR {$k=1, 2, \dots, |\mathcal{A}_i|$}
		\FOR { ${A}_{(i,k)} \in \mathcal{A}_i$}
		\STATE ${L}^{*}_{{A}_{(i,k)}}=\underset{\substack{\mathfrak{C}_z\subseteq {A}_{(i,k)}\\ z \in \left[\binom{c}{r}\right]}}{\text{max}}({L}_{z} )$;
		\FOR {\textbf{all} $\ell\in  [{L}^{*}_{{A}_{(i,k)}}]$ }
		\STATE server transmits $\underset{\substack{z \in \left[\binom{c}{r}\right]\\ \mathfrak{C}_z \subseteq {A}_{(i,k)}}}{\oplus} W^{\textbf{d}(u_{z}(\ell))}_{{A}_{(i,k)}  \backslash \mathfrak{C}_z}$; 
		\ENDFOR
		\ENDFOR
		\ENDFOR
        \ENDFOR
		
		\STATE \textbf{end procedure}
	\end{algorithmic}
	\label{algo2}
 
\end{algorithm} 
We now show that Algorithm \ref{algo2} delivers the demands of all the users. Consider an arbitrary user $u_{a}(\ell)$ demanding a file $W^{{d}(u_{a}(\ell))}$, where $a\in\left[\binom{c}{r}\right]$ and $\ell\in[L_i]$. The user $u_{a}(\ell)$ requires the subfiles $W^{{d}(u_{a}(\ell))}_{\mathcal{D}}$ with $\mathfrak{C}_a \cap \mathcal{D}= \phi$ to fulfill its demands. For such an arbitrary $\mathcal{D}$, the set $\mathcal{S} = \mathcal{D}\cup \mathfrak{C}_a$ is of cardinality greater than or equal to $r$. From Lemmas \ref{occurs} and \ref{occurs_once}, we can say that there exist one $A_{(i,k)}$ which is mapped to $\mathcal{S}$ for some $i\in\left[\binom{c}{r}\right]$ and $k\in \left[|\mathcal{A}_i|\right]$. Accordingly, the server transmits $$\underset{\substack{z \in \left[\binom{c}{r}\right]\\ \mathfrak{C}_z \subseteq \mathcal{S}}}{\oplus} W^{{d}(u_{z}(\ell))}_{A_{(i,k)}  \backslash \mathfrak{C}_z}.$$ As $\mathcal{D}=\mathcal{S}\backslash\mathfrak{C}_a=A_{(i,k)}\backslash\mathfrak{C}_a$, the above transmission can be viewed as $W^{{d}(u_{a}(\ell))}_{\mathcal{D}}$ XORed with other terms, which are available in the caches accessed by $u_{a}(\ell)$. Thus, $u_{a}(\ell)$ is able to decode $W^{{d}(u_{a}(\ell))}_{\mathcal{D}}$ from this transmission. Since, $\mathcal{D}$ is arbitrarily chosen, any file $W^{{d}(u_{a}(\ell))}_{\mathcal{D}}$ with $\mathcal{C}_a \cap \mathcal{D} = \phi$ can be decoded by the user $u_{a}(\ell)$.  Additionally, as $u_{a}(\ell)$ is arbitrarily chosen, the proposed Algorithm \ref{algo2} delivers the demands of all the users. We now compute the per user transmission rate of this delivery algorithm. Note that for arbitrary $i\in \left[\binom{c}{r}\right]$ and $k \in \left[|\mathcal{A}_i|\right]$, the server makes ${L}^*_{A_{(i,k)}}$ transmissions corresponding to a set $A_{(i,k)}$. For each transmission, the number of bits transmitted is $\gamma^{|A_{(i,k)}|-r}(1-\gamma)^{c-|A_{(i,k)}|+r}F$. Consequently, the number of bits transmitted for the total ${L}^*_{A_{(i,k)}}$ transmissions is ${L}^*_{A_{(i,k)}}\gamma^{|A_{(i,k)}|-r}(1-\gamma)^{c-|A_{(i,k)}|+r}F$. As $k$ ranges from $1$ to $|\mathcal{A}_i|$ and $i$ ranges from $1$ to $\binom{c}{r}$, the total bits transmitted is $
 \sum_{i=1}^{\binom{c}{r}}\sum_{k=1}^{|\mathcal{A}_i|}{L}^*_{A_{(i,k)}}\gamma^{|A_{(i,k)}|-r}(1-\gamma)^{c-|A_{(i,k)}|+r}F.$ Upon normalizing over $F$ and $K$ we obtain the per user transmission rate as
\begin{equation}
\label{PerUserTR_Ext}
 \dfrac{\sum_{i=1}^{\binom{c}{r}}\sum_{k=1}^{|\mathcal{A}_i|}{L}^*_{A_{(i,k)}}\gamma^{|A_{(i,k)}|-r}(1-\gamma)^{c-|A_{(i,k)}|+r}}{K}.
\end{equation}

 \subsection{Lower Bound}
 \label{ext_LB}
 Recall that in the proof of Lemma \ref{thm:alpha1}, $\mathcal{Y}(\mathcal{I}_{DM})=\bigcup\limits_{i\in\left[ \binom{c}{r}\right],\ell\in[L] ,\mathcal{S}^{\prime} \in \mathcal{P}_{i} } W^{u_{i}(\ell)}_{\mathcal{S}^{\prime}}$ is a generalized independent set as every subset of $\mathcal{Y}(\mathcal{I}_{DM})$ belongs to the set $\mathcal{J}(\mathcal{I}_{DM})$. By replacing $L$ with $L_i$, $\mathcal{Y}(\mathcal{I}_{DM})$ is modified to $\bigcup\limits_{i\in\left[ \binom{c}{r}\right],\ell\in[L_i] ,\mathcal{S}^{\prime} \in \mathcal{P}_{i} } W^{u_{i}(\ell)}_{\mathcal{S}^{\prime}}$. Following the same logic, to prove that $\bigcup\limits_{i\in\left[ \binom{c}{r}\right],\ell\in[L_i] ,\mathcal{S}^{\prime} \in \mathcal{P}_{i} } W^{u_{i}(\ell)}_{\mathcal{S}^{\prime}}$ is a generalized independent set for the index coding problem considering decentralized multi-access coded caching system with cache access degree $r$, it is sufficient to show that every subset of $\bigcup\limits_{i\in\left[ \binom{c}{r}\right],\ell\in[L_i] ,\mathcal{S}^{\prime} \in \mathcal{P}_{i} } W^{u_{i}(\ell)}_{\mathcal{S}^{\prime}}$ lies in the set $\underset{i \in\left[ \binom{c}{r}\right]}{\cup}\underset{\ell\in [ L_i]}{\cup}\underset{y \in[ F]}{\cup}\left\{\{{d}({u_{i}(\ell,y)})\} \cup B_{u_{i}(\ell,y)}: B_{u_{i}(\ell,y)} \subseteq \mathcal{B}_{u_{i}(\ell,y)}\right\}$, where $\mathcal{B}_{u_{i}(\ell,y)} = [ NF] \backslash\left(\{{d}({u_{i}(\ell,y)})\} \cup \mathcal{X}_{i}\right)$. Taking motivation from the proof of Lemma \ref{thm:alpha1}, consider an arbitrary set of bits  $$D=\left\{W^{u_{z_{1}}(\ell_1)}_ {\mathcal{S}_{1}}, W^{u_{z_{2}}(\ell_2)}_{\mathcal{S}_{2}}, \ldots, W^{u_{z_{J}}(\ell_J)}_{ \mathcal{S}_{J}}\right\} \subseteq \bigcup\limits_{i\in\left[ \binom{c}{r}\right],\ell\in[L_i] ,\mathcal{S}^{\prime} \in \mathcal{P}_{i} } W^{u_{i}(\ell)}_{\mathcal{S}^{\prime}},$$ where ${z}_{1} \leq {z}_{2} \leq \ldots \leq z_{J}$, $\mathcal{S}_{j}\in \mathcal{P}_{z_{j}}$, $z_{j}\in[\binom{c}{r}]$, $\ell_j \in [L_i] $, and $j\in[J]$. Recall that $\mathfrak{C}_{z_1}\cap\mathcal{S}_j=\phi$ as $\mathcal{S}_{j}\in \mathcal{P}_{z_{j}}$ and $z_j\geq z_1$, $\forall j\in[J]$. Therefore, the set of receivers, each demanding a bit in $W^{u_{z_{1}}(\ell_1)}_{ \mathcal{S}_{1}} $, which are associated to the user $u_{z_1}(\ell_1)$ do not have any bits in $D$ as their side information. This implies that $D\in\underset{i \in\left[ \binom{c}{r}\right]}{\cup}\underset{\ell\in [ L_i]}{\cup}\underset{y \in[ F]}{\cup}\left\{\{{d}({u_{i}(\ell,y)})\} \cup B_{u_{i}(\ell,y)}: B_{u_{i}(\ell,y)} \subseteq \mathcal{B}_{u_{i}(\ell,y)}\right\}$. The same can be said for every subset of $\bigcup\limits_{i\in\left[ \binom{c}{r}\right],\ell\in[L_i] ,\mathcal{S}^{\prime} \in \mathcal{P}_{i} } W^{u_{i}(\ell)}_{\mathcal{S}^{\prime}}$ since $D$ is arbitrarily chosen. This proves that $\bigcup\limits_{i\in\left[ \binom{c}{r}\right],\ell\in[L_i] ,\mathcal{S}^{\prime} \in \mathcal{P}_{i} } W^{u_{i}(\ell)}_{\mathcal{S}^{\prime}}$ is a generalized independent set.
 We now compute the size of this set. Consider a arbitrary set $\mathcal{P}_{i}$, for $i\in \left[\binom{c}{r}\right]$. The set $\mathcal{S}^{\prime}$ will correspond to some $k$-th set $P_{(i,k)}\in \mathcal{P}_{i}$. Each $W^{u_{i}(\ell)}_{P_{(i,k)}}$ is of size $\gamma^{|P_{(i,k)}|}(1-\gamma)^{c-|P_{(i,k)}|}F$ bits, where $\ell\in[L_i]$. As $k$ ranges from 1 to $|\mathcal{P}_{i}|$ and $i$ ranges from 1 to $\binom{c}{r}$, we compute $\sum_{i=1}^{\binom{c}{r}}{L}_i\sum_{k=1}^{|\mathcal{P}_i|}\gamma^{|P_{(i,k)}|}(1-\gamma)^{c-|P_{(i,k)}|}F$ as the size of the generalized independent set. Using \eqref{alphaaaa} we can say that the size of the largest generalized independent set forms a lower bound on an optimal linear index code, this implies that 
 \begin{equation}
 \label{mod_alpha}
 \sum_{i=1}^{\binom{c}{r}}{L}_i\sum_{k=1}^{|\mathcal{P}_i|}\gamma^{|P_{(i,k)}|}(1-\gamma)^{c-|P_{(i,k)}|}F.
 \end{equation} 
 is also a lower bound on the optimal linear index code. Recall that the bits transmitted by the server utilizing the delivery scheme of the caching problem serve as an upper bound on optimal linear index code. Therefore, from the proposed per user transmission rate of the decentralized multi-access coded caching scheme with cache access degree $r$ obtained in \eqref{PerUserTR_Ext}, we get an upper bound on the optimal linear index code as 
 \begin{equation}
 \label{kappa_upper_mod}
  \sum_{i=1}^{\binom{c}{r}}{L}^*_{A_{(i,k)}}\sum_{k=1}^{|\mathcal{A}_i|}\gamma^{|A_{(i,k)}|-r}(1-\gamma)^{c-|A_{(i,k)}|+r}F.\end{equation}
 It is to be noted that as $|\mathcal{P}_i|=|\mathcal{A}_i|$ and $|P_{(i,k)}|=|A_{(i,k)}|-r$,  \eqref{mod_alpha} and \eqref{kappa_upper_mod} share almost identical structures, differing only in the terms $L_i$ and ${L}^*_{A_{(i,k)}}$ respectively. Hence, they are not equal. This motivates us to study the conditions for which ${L}^*_{A_{(i,k)}}=L_i$. The following lemma is utilized to find a condition under which ${L}^*_{A_{(i,k)}}=L_i$.
\begin{lemma}
\label{A}
 For any set ${A}_{(i,k)}\in \mathcal{A}_{i}$, every $r$ element subset of $A_{(i,k)}$ will be equal to a set $\mathfrak{C}_z$ with $z \geq i$.
 \end{lemma}
\begin{IEEEproof}
      Any set $A_{(i,k)}$ is of the form $\mathfrak{C}_{i} \cup P_{(i,k)}$ where $P_{(i,k)} \in \mathcal{P}_{i}$. Hence, every element in $A_{(i,k)}$ belongs either to $\mathfrak{C}_{i}$ or to $[c] \backslash [m]$ (using Lemma \ref{C}), where $m = \underset{j\in \mathfrak{C}_{i}}{\max \,}\{j\}.$ Consider any $\mathfrak{C}_z$ which is an $r$ element subset  of $A_{(i,k)}$. Either all the entries of $\mathfrak{C}_z$ will be from $\mathfrak{C}_{i}$ which implies that $z=i$ or there will be at least one element in $\mathfrak{C}_z$ from the set $[c] \backslash [m]$. From the mapping of $\mathfrak{C}_z$ to the binary vectors, we can conclude that $z > i$. This completes the proof.
  \end{IEEEproof}
  \begin{theorem}
  \label{equality}
      If ${L}_1\geq {L}_2 \geq\ldots\geq {L}_{\binom{c}{r}}$, the proposed delivery scheme is optimal and  $$\dfrac{\sum_{i=1}^{\binom{c}{r}}{L}_i\sum_{k=1}^{|\mathcal{A}_i|}\gamma^{|A_{(i,k)}|-r}(1-\gamma)^{c-|A_{(i,k)}|+r}}{K}$$ is the optimal per user transmission rate.
  \end{theorem}
  \begin{IEEEproof}
      Using Lemma \ref{A}, we have ${L}^*_{A_{(i,k)}}=\underset{\substack{\mathfrak{C}_z\subseteq {A}_{(i,k)}\\ z \in \left\{i, i+1,\ldots, \binom{c}{r}\right\}}}{\text{max}}({L}_{z} )$. If ${L}_1\geq {L}_2 \geq\ldots\geq {L}_{\binom{c}{r}}$, we obtain ${L}^*_{A_{(i,k)}}=L_i$. Therefore, we can say that \eqref{mod_alpha} and \eqref{kappa_upper_mod} are equal, implying that the upper and the lower bounds on the optimal linear index code is tight and is equal to $\sum_{i=1}^{\binom{c}{r}}L_i\sum_{k=1}^{|\mathcal{A}_i|}\gamma^{|A_{(i,k)}|-r}(1-\gamma)^{c-|A_{(i,k)}|+r}F$. Upon normalizing over $F$ and $K$ we obtain the optimal per user transmission rate. This proves the optimality of our proposed scheme among all possible linear delivery schemes for the decentralized multi-access coded caching system with cache access degree $r$ having ${L}_1\geq {L}_2 \geq\ldots\geq {L}_{\binom{c}{r}}$.
  \end{IEEEproof} 
\begin{remark}
 Note that the condition ${L}_1\geq {L}_2 \geq\ldots\geq {L}_{\binom{c}{r}}$ imposed on the user-to-cache association profile is sufficient to prove the optimality but is not a necessary condition. Alternate conditions may be explored to establish the optimality of the delivery scheme.
\end{remark}
  
\begin{figure}[h!]
		\centering	
		\includegraphics[scale=0.65]{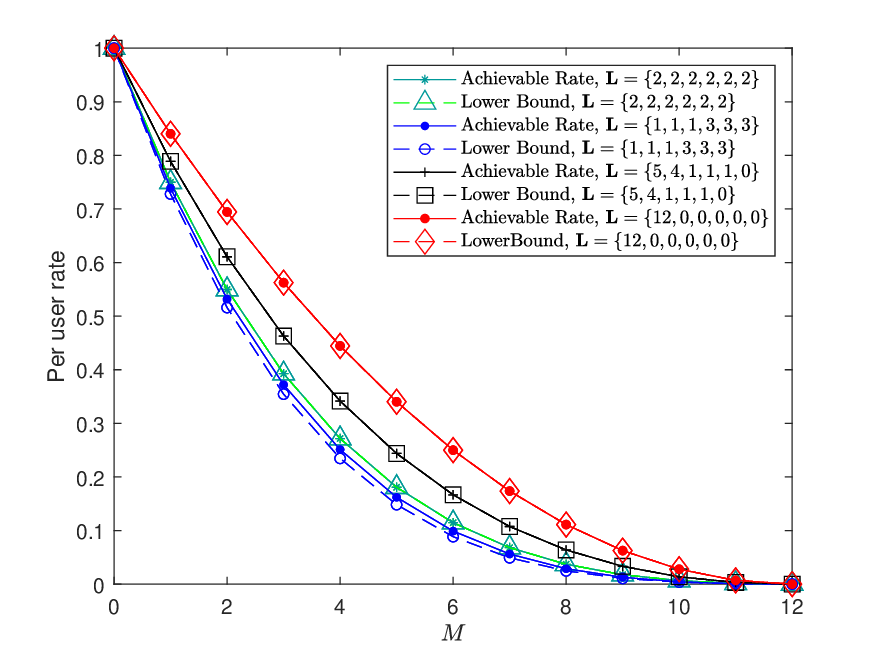}
		\caption{Performance analysis of the generalized decentralized multi-access coded caching problem with $N=K=12$, $c=4$, and $r=2$ for various user-to-cache association profiles.}
		\label{L_comp}
	\end{figure}

 For the decentralized multi-access coded caching system with cache access degree $r$, we verify Theorem \ref{equality} for various user-to-cache association profiles. In Fig. \ref{L_comp}, the per user transmission rates for four different user-to-cache association profiles, along with their corresponding lower bounds, for the parameters $N=K=12$, $c=4$, and $r=2$ are shown. Observe that in Fig. \ref{L_comp}, the per user transmission rate and the corresponding lower bound for $\textbf{L}=(1,1,1,3,3,3)$ do not match as this profile does not satisfy the condition ${L}_1\geq {L}_2 \geq\ldots\geq {L}_{6}$. For the remaining user-to-cache association profiles, which satisfy this condition, the per user transmission rates align tightly with their respective lower bounds.
\section{Conclusion}
\label{conclu}
This paper introduces a system model for a decentralized multi-access coded caching network with cache access degree $r$ and user association $L$. A novel linear delivery scheme is presented for this model and a closed-form expression of per user transmission rate is computed. Using index coding techniques, the proposed scheme is shown to be optimal. Furthermore, it demonstrates that existing frameworks \cite{MaNAccess}, \cite{maddah2014decentralized}, and \cite{MShared} are special cases of the proposed model. Additionally, a more general multi-access coded caching setting is proposed, eliminating all constraints on the number of users and optimal delivery scheme is proposed for the user-to-cache association profiles satisfying ${L}_1\geq {L}_2 \geq\ldots\geq {L}_{\binom{c}{r}}$. Future work may focus on devising an optimal delivery scheme for the generalized decentralized multi-access coded caching problem with arbitrary user-to-cache associations.
\section*{Acknowledgment}
This work was supported by the Science and Engineering Research Board (SERB) of the Department of Science and Technology (DST), Government of India, through J. C. Bose National Fellowship to B. Sundar Rajan.
\balance
\bibliographystyle{IEEEtran}
 
	\bibliography{IEEEabrv,referencesMF}

\begin{thebibliography}{10}
\providecommand{\url}[1]{#1}
\csname url@samestyle\endcsname
\providecommand{\newblock}{\relax}
\providecommand{\bibinfo}[2]{#2}
\providecommand{\BIBentrySTDinterwordspacing}{\spaceskip=0pt\relax}
\providecommand{\BIBentryALTinterwordstretchfactor}{4}
\providecommand{\BIBentryALTinterwordspacing}{\spaceskip=\fontdimen2\font plus
\BIBentryALTinterwordstretchfactor\fontdimen3\font minus
  \fontdimen4\font\relax}
\providecommand{\BIBforeignlanguage}[2]{{%
\expandafter\ifx\csname l@#1\endcsname\relax
\typeout{** WARNING: IEEEtran.bst: No hyphenation pattern has been}%
\typeout{** loaded for the language `#1'. Using the pattern for}%
\typeout{** the default language instead.}%
\else
\language=\csname l@#1\endcsname
\fi
#2}}
\providecommand{\BIBdecl}{\relax}
\BIBdecl

\bibitem{MaN1}
M.~A. Maddah-Ali and U.~Niesen, ``Fundamental limits of caching,'' \emph{IEEE
  Trans. Inf. Theory}, vol.~60, no.~5, pp. 2856--2867, May 2014.

\bibitem{maddah2014decentralized}
M.~A. {Maddah-Ali} and U.~{Niesen}, ``Decentralized coded caching attains
  order-optimal memory-rate tradeoff,'' \emph{IEEE/ACM Trans. Netw.}, vol.~23,
  no.~4, pp. 1029--1040, Aug. 2014.

\bibitem{Combi_multi24}
K.~K.~K. Namboodiri and B.~S. Rajan, ``Combinatorial multi-access coded
  caching: Improved rate-memory trade-off with coded placement,'' \emph{IEEE
  Trans. Inf. Theory}, vol.~70, no.~3, pp. 1787--1805, Mar. 2024.

\bibitem{Combi_multi23}
E.~Peter and B.~S. Rajan, ``Multi-antenna coded caching with combinatorial
  multi-access networks,'' in \emph{Proc. IEEE Int. Symp. Inf. Theory (ISIT)},
  Taipei, Taiwan, Jun. 2023, pp. 126--131.

\bibitem{filesize}
J.~Zhang, X.~Lin, C.~C. Wang, and X.~Wang, ``Coded caching for files with
  distinct file sizes,'' in \emph{Proc. IEEE Int. Symp. Inf. Theory (ISIT)},
  Hong Kong, China, Jun. 2015, pp. 1686--1690.

\bibitem{KNMDHeirarchical}
N.~Karamchandani, U.~Niesen, M.~A. Maddah-Ali, and S.~N. Diggavi,
  ``Hierarchical coded caching,'' \emph{IEEE Trans. Inf. Theory}, vol.~62,
  no.~6, pp. 3212--3229, Jun. 2016.

\bibitem{onlinecaching}
R.~Pedarsani, M.~A. Maddah-Ali, and U.~Niesen, ``Online coded caching,''
  \emph{IEEE/ACM Trans. Netw.}, vol.~24, no.~2, pp. 836--845, Apr. 2016.

\bibitem{D2D}
M.~Ji, G.~Caire, and A.~F. Molisch, ``Fundamental limits of caching in wireless
  {D2D} networks,'' \emph{IEEE Trans. Inf. Theory}, vol.~62, no.~2, pp.
  849--869, Feb. 2016.

\bibitem{nonuniformdemands}
U.~Niesen and M.~A. Maddah-Ali, ``Coded caching with nonuniform demands,''
  \emph{IEEE Trans. Inf. Theory}, vol.~63, no.~2, pp. 1146--1158, Feb. 2017.

\bibitem{Array}
Q.~Yan, M.~Cheng, X.~Tang, and Q.~Chen, ``On the placement delivery array
  design for centralized coded caching scheme,'' \emph{IEEE Trans. Inf.
  Theory}, vol.~63, no.~9, pp. 5821--5833, Sep. 2017.

\bibitem{WUCCMF}
Y.-P. Wei and S.~Ulukus, ``Coded caching with multiple file requests,'' in
  \emph{Proc. 55th Annu. Allerton Conf. Commun. Control Comput. (Allerton)},
  Monticello, USA, Oct. 2017, pp. 437--442.

\bibitem{shared}
E.~{Parrinello}, A.~{Ünsal}, and P.~{Elia}, ``Fundamental limits of coded
  caching with multiple antennas, shared caches and uncoded prefetching,''
  \emph{IEEE Trans. Inf. Theory}, vol.~66, no.~4, pp. 2252--2268, Apr. 2020.

\bibitem{Main_Access}
J.~Hachem, N.~Karamchandani, and S.~N. Diggavi, ``Coded caching for multi-level
  popularity and access,'' \emph{IEEE Trans. Inf. Theory}, vol.~63, no.~5, pp.
  3108--3141, May 2017.

\bibitem{MAGains}
B.~Serbetci, E.~Parrinello, and P.~Elia, ``Multi-access coded caching: Gains
  beyond cache-redundancy,'' in \emph{Proc. IEEE Inf. Theory Workshop (ITW)},
  Visby, Sweden, Aug. 2019, pp. 1--5.

\bibitem{firstmaJ}
K.~S. Reddy and N.~Karamchandani, ``Rate-memory trade-off for multi-access
  coded caching with uncoded placement,'' \emph{IEEE Trans. Commun.}, vol.~68,
  no.~6, pp. 3261--3274, Jun. 2020.

\bibitem{SasiImproved}
S.~Sasi and B.~S. Rajan, ``Multi-access coded caching scheme with linear
  sub-packetization using {PDAs},'' in \emph{Proc. IEEE Int. Symp. Inf. Theory
  (ISIT)}, Melbourne, Australia, Jul. 2021, pp. 861--866.

\bibitem{CLWZC}
M.~Cheng, D.~Liang, K.~Wan, M.~Zhang, and G.~Caire, ``A novel transformation
  approach of shared-link coded caching schemes for multiaccess networks,'' in
  \emph{Proc. IEEE Int. Symp. Inf. Theory (ISIT)}, Melbourne, Australia, Jul.
  2021, pp. 849--854.

\bibitem{MaNAccess}
P.~N. Muralidhar, D.~Katyal, and B.~S. Rajan, ``Maddah-ali-niesen scheme for
  multi-access coded caching,'' in \emph{Proc. IEEE Information Theory Workshop
  (ITW)}, Kanazawa, Japan, Oct. 2021, pp. 1--6.

\bibitem{MALowerB}
F.~Brunero and P.~Elia, ``The exact load-memory tradeoff of multi-access coded
  caching with combinatorial topology,'' in \emph{Proc. IEEE Int. Symp. Inf
  Theory (ISIT)}, Espoo, Finland, Jun. 2022, pp. 1701--1706.

\bibitem{Combi}
------, ``Fundamental limits of combinatorial multi-access caching,''
  \emph{IEEE Trans. Inf. Theory}, vol.~69, no.~2, pp. 1037--1056, Feb. 2023.

\bibitem{Conf_PMA}
P.~Trinadh, M.~Dutta, A.~Thomas, and B.~S. Rajan, ``Decentralized multi-access
  coded caching with uncoded prefetching,'' in \emph{Proc. IEEE Inf. Theory
  Workshop (ITW)}, Kanazawa, Japan, Nov. 2021, pp. 1--6.

\bibitem{MShared}
M.~Dutta and A.~Thomas, ``Decentralized coded caching for shared caches,''
  \emph{IEEE Commun. Lett.}, vol.~25, no.~5, pp. 1458--1462, May 2021.

\bibitem{KVRDecentralized}
N.~S. Karat, K.~L.~V. Bhargav, and B.~S. Rajan, ``On the optimality of two
  decentralized coded caching schemes with and without error correction,'' in
  \emph{Proc. IEEE Int. Symp. Inf. Theory (ISIT)}, Los Angeles, USA, Jun. 2020,
  pp. 1664--1669.

\bibitem{ECSBP}
N.~S. Karat, A.~Thomas, and B.~S. Rajan, ``Error correction in coded caching
  with symmetric batch prefetching,'' \emph{IEEE Trans. Commun.}, vol.~67,
  no.~8, pp. 5264--5274, Aug. 2019.

\bibitem{Katyal}
D.~Katyal, P.~N. Muralidhar, and B.~S. Rajan, ``Multi-access coded caching
  schemes from cross resolvable designs,'' \emph{IEEE Trans. Commun.}, vol.~69,
  no.~5, pp. 2997--3010, May 2021.

\bibitem{MA_PIR}
K.~Vaidya and B.~Sundar~Rajan, ``Multi-access cache-aided multi-user private
  information retrieval,'' \emph{IEEE Trans. Commun.}, early access, Mar. 11,
  2024, doi: 10.1109/TCOMM.2024.33758.

\bibitem{index}
Y.~Birk and T.~Kol, ``Coding on demand by an informed source {(ISCOD)} for
  efficient broadcast of different supplemental data to caching clients,''
  \emph{IEEE Trans. Inf. Theory}, vol.~52, no.~6, pp. 2825--2830, Jun. 2006.

\bibitem{DSC}
S.~H. Dau, V.~Skachek, and Y.~M. Chee, ``Error correction for index coding with
  side information,'' \emph{IEEE Trans. Inf. Theory}, vol.~59, no.~3, pp.
  1517--1531, Mar. 2013.

\bibitem{firstma}
K.~S. {Reddy} and N.~{Karamchandani}, ``On the exact rate-memory trade-off for
  multi-access coded caching with uncoded placement,'' in \emph{Proc. Int.
  Conf. Signal Process. Commun. (SPCOM)}, Bangalore, India, Jul. 2018, pp.
  1--5.

\bibitem{KSTR}
N.~S. {Karat}, S.~{Dey}, A.~{Thomas}, and B.~S. {Rajan}, ``An optimal linear
  error correcting delivery scheme for coded caching with shared caches,'' in
  \emph{Proc. IEEE Int. Symp. Inf. Theory (ISIT)}, Paris, France, Jul. 2019,
  pp. 1217--1221.

\end{thebibliography}

\end{document}